\documentclass{aa}
\usepackage{natbib}
\bibpunct{(}{)}{;}{a}{}{,}

\usepackage{graphicx}
\usepackage[varg]{txfonts}

\def\totdif#1#2{\frac{{\rm d}#1}{{\rm d}#2}}

\begin{document} 

\title{Calibration of the mixing-length parameter $\alpha$ \\
  for the MLT and FST models by matching with CO$^5$BOLD models}
\subtitle{}

\author{T. Sonoi\inst{1,2}
  \and
  H.-G. Ludwig\inst{3,4}
  \and
  M.-A. Dupret\inst{5}
  \and
  J. Montalb\'an\inst{6}
  \and
  R. Samadi\inst{2}
  \and
  K. Belkacem\inst{2}
  \and\\
  E. Caffau\inst{4}
  \and
  M.-J. Goupil\inst{2}
}

\institute{Astronomical Institute, Tohoku University, 6-3 Aramaki Aza-Aoba, Aoba-ku Sendai, 980-8578, Japan
  \and
  LESIA, Observatoire de Paris,
  {PSL University,}
  CNRS, Universit\'e Pierre et Marie Curie, Universit\'e Denis Diderot, F-92195 Meudon, France
  \and
  Zentrum f\"ur Astronomie der Universit\"at Heidelberg, Landessternwarte, K\"onigstuhl 12, D-69117 Heidelberg, Germany
  \and
  GEPI, Observatoire de Paris, 
  {PSL University,} 
  CNRS, Place Jules Janssen, F-92190 Meudon, France
  \and
  Institut d'Astrophysique et de G\'eophysique, Universit\'e de Li\`ege, All\'ee du 6 Ao\^ut 17, 4000, Li\`ege, Belgium
  \and
  Dipartimento di Fisica e Astronomia, University of Padova, Vicolo dell'Osservatorio 3, I-35122 Padova, Italy
}

\offprints{\tt sonoi@astr.tohoku.ac.jp}
\date{\today}

 
\abstract
    {Space observations by the CoRoT and {\it Kepler} missions provided a wealth of high-quality seismic data for a large number of stars from the main sequence to the red-giant phases. One main goal of these missions is to take advantage of the rich spectra of solar-like oscillations to perform precise determinations of stellar characteristic parameters. To make the best of such data, we need theoretical stellar models with precise near-surface structure, which has significant influence on solar-like oscillation frequencies. The mixing-length parameter is a key factor to determine the near-surface structure of stellar models. In current versions of the convection formulations used in stellar evolution codes, the mixing-length parameter is a free parameter that needs to be properly specified.}
    {We aim at determining appropriate values of the mixing-length parameter, $\alpha$, to be used consistently with the adopted convection formulation when computing stellar evolution models across the Hertzsprung-Russell diagram. This determination is based on 3D hydrodynamical simulation models.}
    {We calibrated $\alpha$ values by matching entropy profiles of 1D envelope models with those of hydrodynamical 3D models of solar-like stars produced by the CO$^5$BOLD code. For such calibration, previous works concentrated on the classical mixing-length theory (MLT). Here we also analyzed the full spectrum turbulence (FST) models. For construction of the atmosphere part in the 1D models, we use the Eddington grey $T(\tau)$ relation and the one with the solar-calibrated Hopf-like function.}
    {For both the MLT and FST models with a mixing length $l=\alpha H_p$, calibrated $\alpha$ values increase with increasing surface gravity or decreasing effective temperature. For the FST models, we carried out an additional calibration using an $\alpha^*$ value defined as $l=r_{\rm top}-r+\alpha^*H_{p,{\rm top}}$, and $\alpha^*$ is found to increase with surface gravity and effective temperature. We provide tables of the calibrated $\alpha$ values across the $T_{\rm eff}$--$\log\,g$ plane for the solar metallicity. By computing stellar evolution with varying $\alpha$ based on our 3D $\alpha$ calibration, we find that the change from the solar $\alpha$ to the varying $\alpha$ shifts evolutionary tracks particularly for the FST model. As for the correspondence to the 3D models, the solar Hopf-like function generally gives a photospheric-minimum entropy closer to a 3D model than the Eddington $T(\tau)$. The structure below the photosphere depends on the adopted convection model. However, we cannot obtain a definitive conclusion about which convection model gives the best correspondence to the 3D models. This is because each 1D physical quantity is related via an EOS, but it is not the case for the averaged 3D quantities. Although the FST models with $l=r_{\rm top}-r+\alpha^*H_{p,{\rm top}}$ are found to give the oscillation frequencies closest to the solar observed frequencies, their acoustic cavities are formed with compensatory effects between deviating density and temperature profiles near the top of the convective envelope. In future work, an appropriate treatment of the top part of the 1D convective envelope is necessary, for example, by considering turbulent pressure and overshooting.}
    {}

\keywords{Convection - Stars:late-type - Stars: solar-type}

\titlerunning{Calibration of $\alpha$ for MLT and FST convection models}
\maketitle
%

\section{Introduction}
High-precision photometry obtained by the  space missions CoRoT and {\it Kepler}/K2 provided a wealth of high-quality data which enable us to carry out asteroseismic studies with unprecedented quality. The PLATO spacecraft is planned to be launched as an ESA's mission to perform precise photometry for detection of exoplanet transits as well as variations in stellar brightness. It is expected to provide a wealth of high-quality seismic data  for an even larger set of bright stars. Particularly, rich spectra of solar-like oscillations can lead to tight constraints on stellar global parameters such as age, mass and radius and properties of the interior structure. However, this requires theoretical stellar models that have comparable quality than the observations.

Today, the so-called surface effects  prevent  us from obtaining correct theoretical frequencies for solar-like oscillations. These effects originate   from an incorrect modeling of the near-surface structure mainly due to a too approximate determination of the superadiabatic temperature  gradient. The mixing-length theory (MLT) proposed by \citeauthor{BV58} (\citeyear{BV58}, hereafter BV) has been commonly adopted to model stellar convection in stellar evolutionary codes. The assumption of a single size of convection eddies is inconsistent with near-surface turbulence. In addition, it is necessary to give values of several free parameters included in the adopted convection formulation. The most prominent free parameter is the mixing length, which is defined as $l=\alpha H_p$, where $H_p$ is the pressure scale height, and $\alpha$ a free parameter. We suffer from the arbitrariness of the $\alpha$ value, and a constraint based on physics is required.

On the other hand, radiation-coupled hydrodynamical (RHD) simulations succeeded in producing realistic turbulence profiles in the near-surface regions. Using results of such simulations, \cite{Steffen93} performed a calibration of the $\alpha$ value using a 2D RHD model of the Sun. Ludwig, Freytag \& Steffen (\citeyear{Ludwig99}, hereafter LFS) and \cite{Freytag99} extended the calibration to different solar-like stars. \cite{Trampedach97} performed the calibration with 3D RHD models. Their work was recently updated by \cite{Trampedach14b} using an increased number of 3D models introduced by \cite{Trampedach13}, who adopted the \cite{Stein98} code. \cite{Ludwig08} adopted 3D models produced by the CO$^5$BOLD code \citep{Freytag02,Wedemeyer04,Freytag12}, and \cite{Magic15} followed with 3D models by the STAGGER code \citep{Magic13}. We note that both the \cite{Stein98} code and the STAGGER code originate from a code developed by \cite{Galsgaard96} \citep{Kritsuk11}. Compared to the 2D models, the 3D models are likely to result in higher $\alpha$ values due to more efficient convection. The calibrations with RHD models have shown that the MLT $\alpha$ value decreases with increasing effective temperature or decreasing surface gravity. \cite{Magic15} also confirmed that it generally decreases with increasing metallicity. In such calibrations, different treatments of atmosphere parts in calibrated 1D models have led to quite a bit of conflicting results among different previous works. Nevertheless, such calibrations would give us physically guaranteed values of $\alpha$. 

For 1D evolutionary computations, the full spectrum turbulence (FST) models were newly proposed by \citeauthor{CM91} (\citeyear{CM91}, hereafter CM) and Canuto, Goldman \& Mazzitelli (\citeyear{CGM96}, hereafter CGM). While the MLT assumes a single size of convection eddies, CM and CGM models take into account contribution of different eddy sizes based on the Eddy Damped Quasi-Normal Markovian model. In spite of the improvement in describing convective fluxes, CM and CGM use an incompressible turbulence model. As a consequence, a distance scale does not appear naturally. Besides, even if one chooses the distance to the convection boundary as adopted in the CM paper, it is not able to fit the solar radius with the accuracy required for seismic studies. Therefore, an ad-hoc free parameter for the scale distance must be introduced to compensate for other uncertainties in stellar physics and get such precision. So far the calibration with the FST model has been performed only with 2D models (LFS; \citeauthor{Freytag99} \citeyear{Freytag99}). As it has been shown that the $\alpha$ value is so different between 2D and 3D models for MLT, it is worth investigating the $\alpha$ values of CM and CGM also for 3D models. 

In this work, we performed the calibration by matching standard 1D envelope models with 3D RHD models produced by the CO$^5$BOLD code. The number of the models was increased compared to \cite{Ludwig08}, and a wide range in the $T_{\rm eff}-g$ plane is covered. We investigated $\alpha$ values adopting the CM and CGM models as well as MLT in the 1D models. In Sect. \ref{sec:models}, we present the 1D and 3D models that we adopt. In Sect. \ref{sec:calib}, we introduce our method for the calibration and discuss the results. Discussion and conclusion are given in Sects. \ref{sec:discussion} and \ref{sec:conclusion}, respectively.

\section{Theoretical models}
\label{sec:models}
\subsection{3D atmosphere models}

We adopted 25 3D-RHD models with the solar metallicity from the CIFIST grid \citep{Ludwig09}, which was produced by the CO$^5$BOLD code \citep{Freytag02,Wedemeyer04,Freytag12}. This code solves the time-dependent equations of compressible hydrodynamics coupled to radiative transfer in a constant gravity field in a Cartesian computational domain which is representative of a volume located at the stellar surface. The equation of state takes into account the ionization of hydrogen and helium, as well as the formation of H$_2$ molecules according to Saha-Boltzmann statistics following \cite{Wolf83}. We used the multi-group opacities based on monochromatic opacities stemming from the MARCS stellar atmosphere package \citep{Gustafsson08}. For the calculation of the opacity using the MARCS package, the chemical abundance basically follows the solar mixture by \cite{Grevesse98}. For the CNO elements, we adopted \cite{Asplund05}.

In the RHD simulation, a specific entropy value is given to ascending convective flows entering from the bottom of a simulation box. We define this entropy value as $s_{\rm 3D,bot}$. Namely, $s_{\rm 3D,bot}$ is an input parameter, while the effective temperature is not. Since the ascending flows are adiabatic, their entropy values are determined by $s_{\rm 3D,bot}$ until reaching the vicinity of the photosphere. By its nature, $s_{\rm 3D,bot}$ can be assumed to equal the asymptotic entropy value in the adiabatic limit of the bottom of a convective envelope, $s_{\rm asy}$. And it was used as a constraint for the $\alpha$ calibration \citep{Steffen93}.

On the other hand, descending flows experience significant heat exchange in the superadiabatic layers just below the photosphere, and then their entropy values are certainly lower than the asymptotic value in the bottom part of the convective envelope, and they reduce the horizontally averaged value. Some models are not deep enough for their averaged entropy values to reach the asymptotic values at the bottoms of the simulation boxes. Thus, it would be preferable to use the asymptotic value for constraining 1D envelope models.

\subsection{1D envelope models}
\label{sec:1D}

Calibration of the mixing-length parameter $\alpha$ for the 1D convection models was performed by matching the entropy value at the adiabatic bottom part of convective envelopes in 1D models with the asymptotic entropy of the 3D models. To do so, we constructed 1D envelope models using a shooting code developed by LFS, which performs integration from atmosphere surface without boundary conditions imposed at a stellar center. Same as the CO$^5$BOLD code, the equation of state follows \cite{Wolf83}, and the opacity is based on the MARCS package \citep{Gustafsson08}. Although the MARCS package provides opacity for different wavelength bands, which have been adopted to the CO$^5$BOLD 3D simulations, a grey Rosseland-mean version was adopted to the 1D models, while the same chemical abundance as the 3D models was adopted. In this work, we did not include turbulent pressure in the 1D models following common practice in stellar structure models. Hence, our calibrated $\alpha$ values should be adopted to evolutionary computation without turbulent pressure, while they include information of turbulent pressure in the 3D models. 

In order to model convection in 1D envelope models, we adopted the convective fluxes from CM and CGM formulations, and those from MLT for comparison with them and with previous works. For MLT, we adopted two formalisms proposed by BV and Henyey, Vardya \& Bodenheimer (\citeyear{Henyey65}, hereafter HVB). The BV version assumes a linear temperature distribution inside a convective bubble, while the HVB version satisfies a diffusion equation for $T^4$ with a constant diffusion parameter. In previous $\alpha$ calibrations, the BV version was adopted by LFS and \cite{Trampedach13}, while the HVB version by \cite{Magic15}.

We need to evaluate the radiative temperature gradient $\nabla_{\rm rad}$ for use of the Schwarzschild criterion and the convection models in a convective region. In a radiative region, the radiative temperature gradient represents the actual temperature gradient. We define it as
\begin{eqnarray}
  \nabla_{\rm rad}\equiv\left(\totdif{\ln T}{\ln p_{\rm g}}\right)_{\rm rad}=\frac{3}{16\sigma}\frac{\kappa F_{\rm tot} p_{\rm g}}{g T^4}\left(1+\totdif{q(\tau)}{\tau}\right),
  \label{eq:nabla_rad}
\end{eqnarray}
where $\sigma$ is the Stefan-Boltzmann constant, $\kappa$ the Rosseland-mean opacity per unit mass, $F_{\rm tot}$ total (radiative plus convective) flux, $p_{\rm g}$ gas pressure, $g$ gravitational acceleration, $T$ temperature. $\tau$ is the optical depth, determined by ${\rm d}\tau\equiv-\kappa\rho{\rm d}r$, where $r$ is radius and $\rho$ density. $q(\tau)$ is a Hopf function, given by a $T(\tau)$ (temperature -- optical depth) relation for radiative equilibrium. Equation (\ref{eq:nabla_rad}) satisfies the differentiation of
\begin{eqnarray}
  \frac{4}{3}\left(\frac{T}{T_{\rm eff}}\right)^4=\tau+q(\tau),
  \label{eq:Ttau}
\end{eqnarray}
where $T_{\rm eff}$ is the effective temperature, in the optically thin limit. Toward optically thicker layers, Eq. (\ref{eq:nabla_rad}) becomes close to the diffusion approximation since ${\rm d}q/{\rm d}\tau\rightarrow 0$ ($\tau\rightarrow\infty$).

The choice of a $T(\tau)$ relation significantly affects an entropy profile. In previous $\alpha$ calibrations, LFS and \cite{Trampedach14b} adopted $T(\tau)$ relations derived from their RHD models. On the other hand, \cite{Magic15} adopted a 1D atmosphere code \citep{Magic13} which solves radiative transfer by itself.

In this work, we adopted fixed $T(\tau)$ relations which can be easily implemented into 1D stellar evolution codes. The first one is the Eddington grey $T(\tau)$ relation ($q(\tau)=2/3$), which has been commonly implemented in many stellar evolution codes. Another one is a $T(\tau)$ relation with the solar-calibrated Hopf-like function (J. Christensen-Dalsgaard, private communication),
\begin{eqnarray}
  q(\tau)=1.036-0.3134\exp(-2.448\tau)-0.2959\exp(-30.0\tau),
  \label{eq:valc}
\end{eqnarray}
which is based on Model C of \citeauthor{Vernazza81} (\citeyear{Vernazza81}, VAL-C).
Model C was derived by solving non-LTE radiative transfer, statistical equilibrium of atomic levels and hydrostatic equilibrium so that a spectrum matches with the one obtained by the {\it Skylab} observation of the quiet Sun. As shown below, this Hopf-like function actually gives much better correspondence to the 3D CO$^5$BOLD models than the Eddington $T(\tau)$. \cite{Salaris15} showed that whether to use a $T(\tau)$ relation based on RHD simulation would have a comparable effect on evolutionary tracks with which to use a solar $\alpha$ or varying $\alpha$ based on calibration with RHD models. However, we note that the use of an RHD $T(\tau)$ relation does not necessarily reproduce a structure identical to original 3D atmosphere due to the different treatment of the radiative transfer between 1D and 3D models \citep{Ludwig08}.

For the integration of 1D envelopes, in addition to the chemical abundance, we need three global parameters, effective temperature $T_{\rm eff}$, surface gravity acceleration $g$ and stellar total mass $M$. Although 3D values were used for $T_{\rm eff}$ and $g$, $M$ was set to $1M_\odot$ for all the models, which is justified by small sensitivity of $\alpha$ to $M$.

\begin{figure}
  \centering
  \includegraphics[width=\hsize]{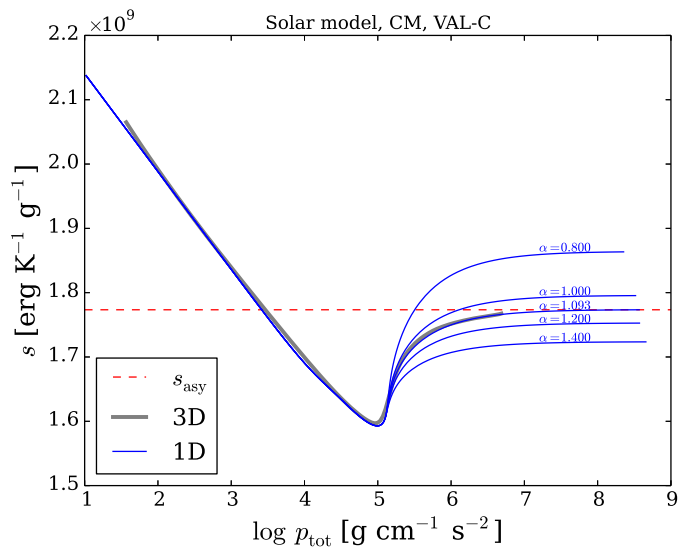}
  \caption{Entropy as function of total pressure for the solar model. The thick grey line is the temporally and horizontally averaged entropy in the 3D model, and the thin blue lines are those in the 1D models with different values of the mixing-length parameter $\alpha$, which are indicated near the blue lines. For the 1D models shown here, CM and the VAL-C $T(\tau)$ relation were used. The horizontal dashed red line indicates the asymptotic entropy of the 3D model, $s_{\rm asy}$.}
  \label{fig:diffalpha}
\end{figure}

\section{Calibration of mixing-length parameter $\alpha$}
\label{sec:calib}

\subsection{Method}

We calibrated values of the mixing-length parameter $\alpha$ for the 25 3D models with the solar metallicity. Figure \ref{fig:diffalpha} shows the calibration for the solar 3D model adopting the CM convection model and VAL-C $T(\tau)$ relation to the corresponding 1D model. The entropy decreases toward the interior in the radiative part of the atmosphere, while it increases in the convective region, which begins near the photosphere. While the grey thick line is the temporally and horizontally averaged profile of the 3D model, the blue thin lines are those of the 1D model with different $\alpha$ values. 

As $\alpha$ increases, the entropy jump between the top of the convective envelope and the adiabatic interior becomes smaller.
It corresponds to a tendency that convection becomes more efficient so that the temperature gradient in superadiabatic layers decreases.
By a bisection method, we searched for an appropriate $\alpha$, which makes the entropy value at the 1D bottom close to the asymptotic entropy, $s_{\rm asy}$, indicated as the horizontal dashed red line. 

Since the MARCS opacity table ranges up to $\log\,T=4.5$ and $\log\,p_{\rm g}=9.0$, we cannot fully complete the 1D integration down to the bottom of a convective envelope. However, an entropy value close to the one at the bottom can still be obtained since it is almost constant in an adiabatic bottom part of a convective envelope. In this work, we stopped the integration when either $T$ or $p_{\rm g}$ reached its upper limit in the opacity table. However, if $\nabla-\nabla_{\rm ad}>10^{-3}$, we proceeded further until $\nabla-\nabla_{\rm ad}$ was reduced to $10^{-3}$. In this case, the opacity was evaluated by setting $T$ or $p_{\rm g}$ to the upper limit value in the table, which causes underestimate of an opacity value. It is used to evaluate radiative temperature gradient $\nabla_{\rm rad}$. Although $\nabla_{\rm rad}\propto\kappa p_{\rm g}/T^4$ in an optically thick case, its variation with depth depends on that of the pressure even more than on the opacity. In such a situation, another possibility might be use of another opacity table having data for high temperature, for instance. However, it could cause loss of smoothness and inconsistency with the upper layers constructed by the MARCS opacity.

\begin{table}
  \centering
  \caption{Calibrated $\alpha$ values for the solar model.}
  \label{tab:sol}
  \begin{tabular}{lcc}\hline\hline
            & VAL-C & Eddington \\ \hline
    CM      & 1.093 & 0.933 \\
    CGM     & 0.821 & 0.702 \\
    CM2     & 0.731 & 0.533 \\
    CGM2    & 0.355 & 0.216 \\
    MLT(BV) & 2.051 & 1.780 \\
    MLT(HVB)& 2.261 & 1.956 \\
    \hline
  \end{tabular}
\end{table}

\subsection{Solar $\alpha$ values}
\label{sec:asol}

Table \ref{tab:sol} summarizes the calibrated $\alpha$ values for the solar model. $\alpha$ has a different meaning among the frameworks of the different convection models. In CM and CGM, mixing length is defined as $l=\alpha H_p$. However, unlike for MLT, this $\alpha$ value is smaller than one \citep[see also][]{CGM96}. CM2 and CGM2 use the mixing length introduced in the CGM paper: $l=r_{\rm top}-r+\alpha^* H_{p,{\rm top}}$. The subscript ``top'' refers to values at the top of the convective envelope. The last term introduces a correction with respect to the mixing length used in the CM paper ($z$: distance to the convection boundary), and it can be interpreted as a kind of overshooting above the convection boundary. The corresponding parameter $\alpha^*$ is expected hence to be much smaller than the $\alpha$ values appearing in the other convection models. Moreover, the effect on the temperature gradient of varying $\alpha^*$ is much smaller than that of varying $\alpha$ in CM, CGM and MLT.

The comparisons of the $\alpha$ values with the previous calibrations by matching with RHD models (\citeauthor{Magic15} \citeyear{Magic15}; \citeauthor{Trampedach14b} \citeyear{Trampedach14b}; LFS) are discussed in Sects. \ref{sec:Magic15} to \ref{sec:LFS}. On the other hand, the $\alpha$ values have been calibrated also empirically by matching observational stellar properties. As a previous work adopting the Eddington $T(\tau)$ relation, \cite{Samadi06} reported $\alpha=0.69$ for CGM and $\alpha=1.76$ for MLT(BV). Our calibrated $\alpha$ values (0.702 and 1.780 for CGM and MLT(BV), respectively), shown in Table \ref{tab:sol}, are in good agreement with them.

We also tested the solar $\alpha$ values using the stellar evolution code ATON 3.1 \citep{Ventura08}. We computed 1$M_\odot$ models using three different implementations of convection (MLT-BV, CGM and CGM2). We adopted the \cite{GN93}'s solar composition, no microscopic diffusion. We derived the values of the different mixing-length parameters with the constraint of the solar radius at the solar luminosity, obtaining: $\alpha=1.784$ for MLT(BV), 0.6907 and 0.1982 for CGM and CGM2 respectively. These $\alpha$ values are in good agreement with the calibration by the envelope models with the Eddington $T(\tau)$ (the right column of Table \ref{tab:sol}), although it is not the $T(\tau)$ relation used for the ATON models.

\cite{Salaris15} calibrated not only for the Eddington $T(\tau)$ but also for VAL-C with MLT(BV). They reported $\alpha=1.69$ for the Eddington $T(\tau)$, and 1.90 for VAL-C. Although these $\alpha$ values are $\sim 0.1$ smaller than our values, they showed that VAL-C gives a certainly larger $\alpha$ value than the Eddington $T(\tau)$ similarly to our result.

\begin{figure*}
  \begin{minipage}{0.49\textwidth}
    \centering
    \includegraphics[width=\hsize]{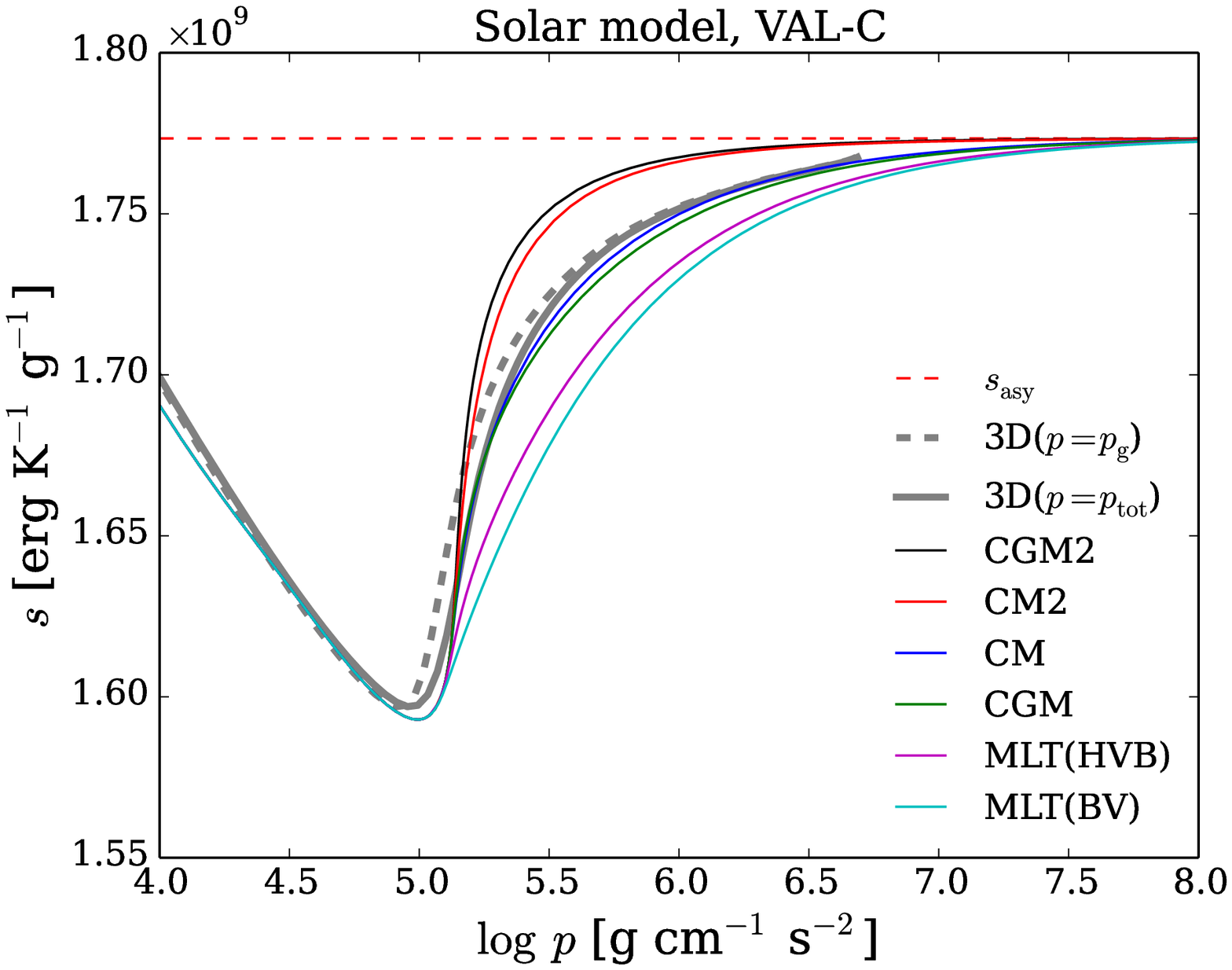}
  \end{minipage}
  \begin{minipage}{0.49\textwidth}
    \centering
    \includegraphics[width=\hsize]{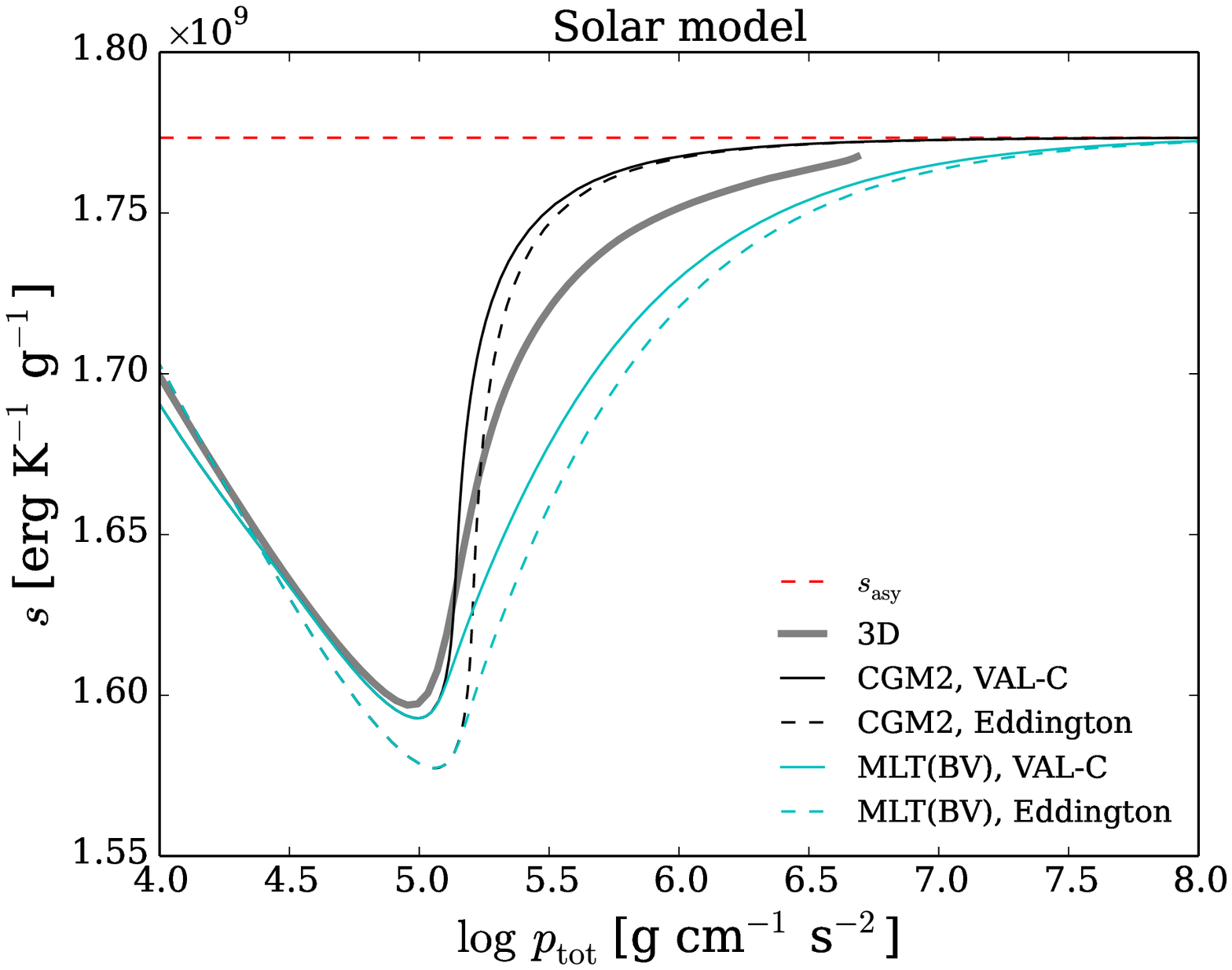}
  \end{minipage}
  \caption{Same as Fig. \ref{fig:diffalpha}, but for calibrated 1D models with different convection models and $T(\tau)$ relations. While the mixing length is determined as $l=r_{\rm top}-r+\alpha^* H_{p,{\rm top}}$, where the subscript ``top'' means the top of the convective envelope, for CM2 and CGM2, $l=\alpha H_p$ for the other cases. In the left panel, all the 1D models were constructed using the VAL-C $T(\tau)$. The right panel shows 1D models with CGM2 and the BV version of MLT both for the VAL-C (the solid lines) and the Eddington $T(\tau)$ (the dashed lines). The 3D profile is shown as function of total pressure (the solid grey line). In the left panel, it is also shown as function of gas pressure (the dashed grey line).}
  \label{fig:diffconv}
\end{figure*}

\begin{figure*}
  \begin{minipage}{0.49\textwidth}
    \centering
    \includegraphics[width=\hsize]{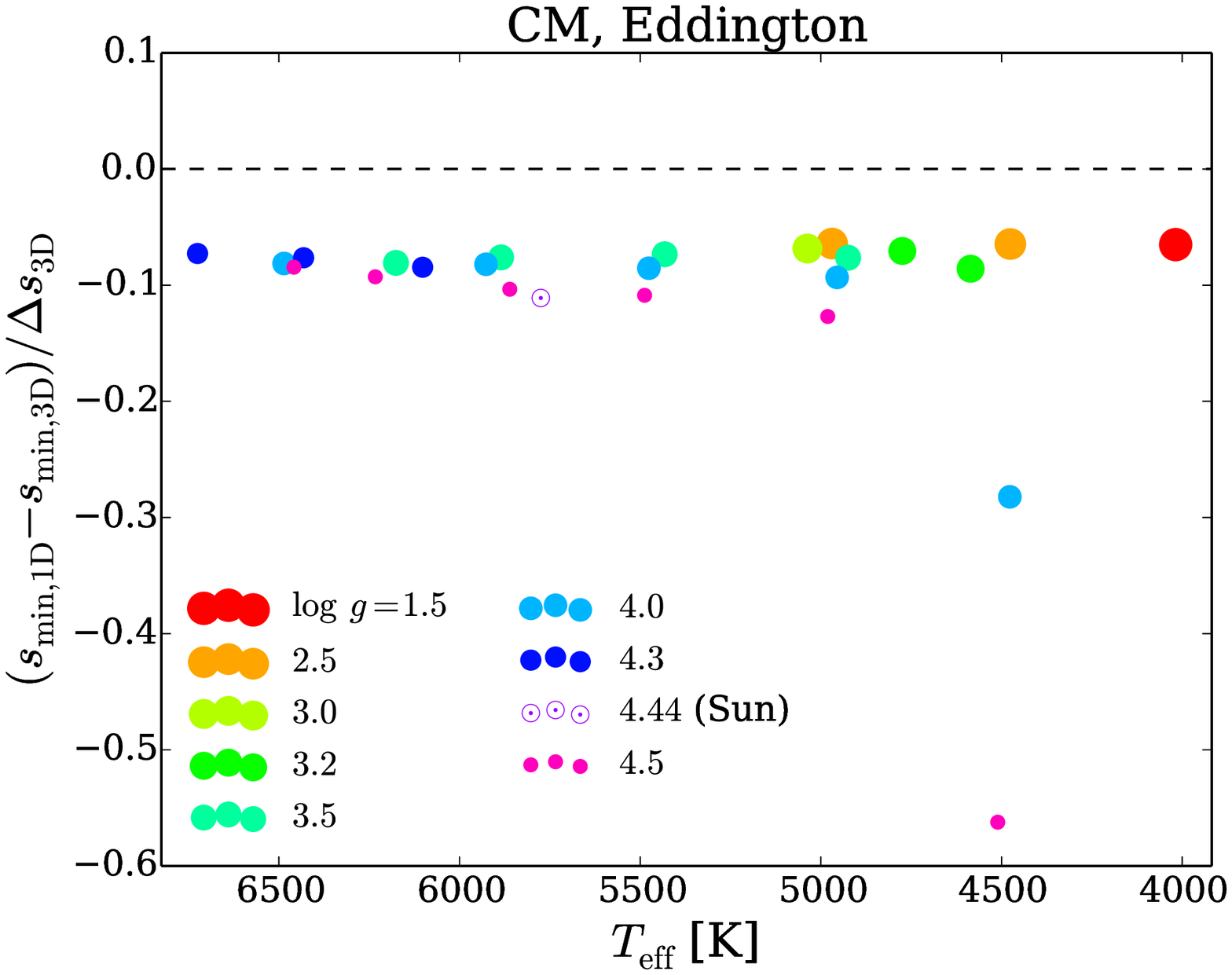}
  \end{minipage}
  \begin{minipage}{0.49\textwidth}
    \centering
    \includegraphics[width=\hsize]{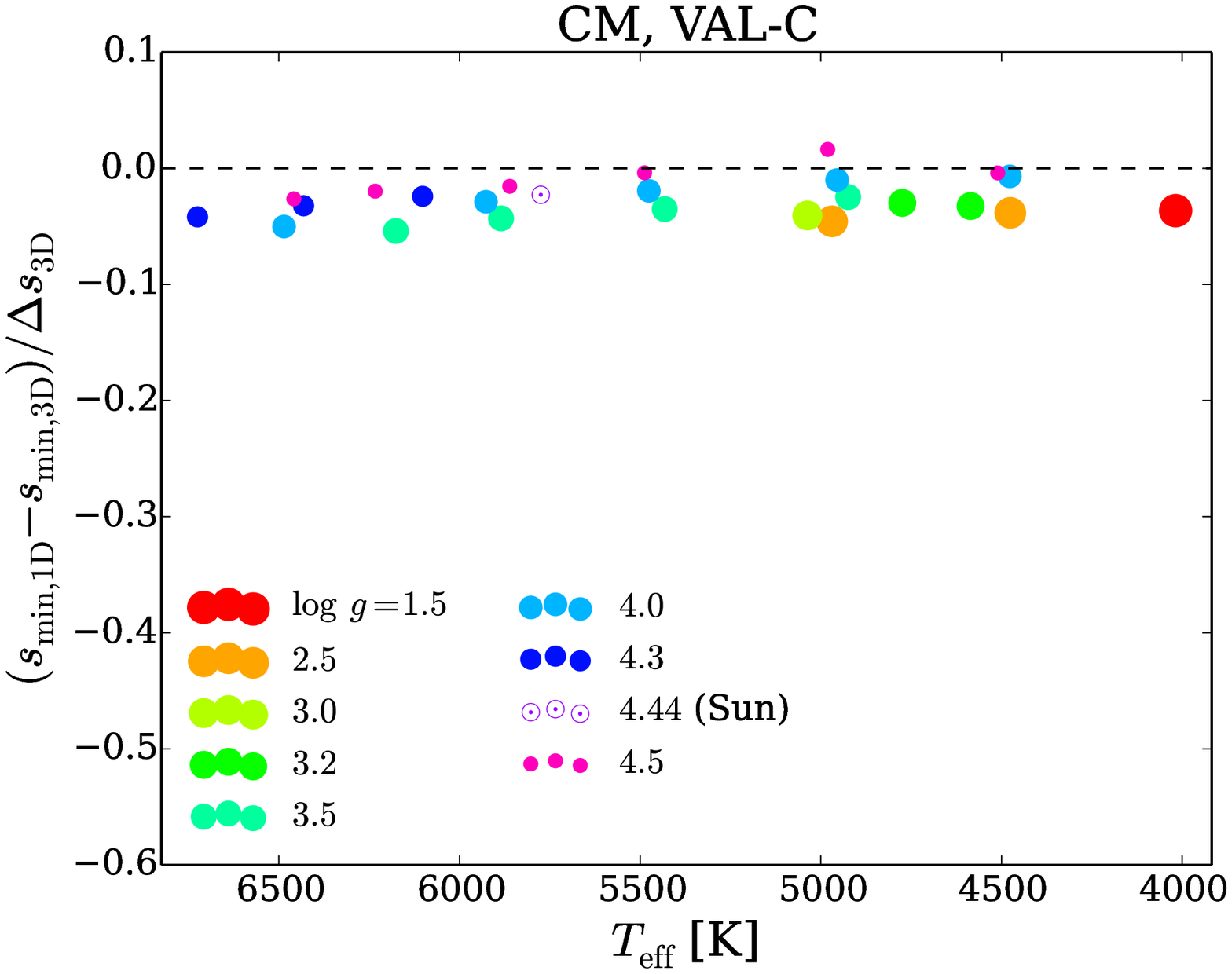}
  \end{minipage}
  \caption{Deviation of 1D photospheric-minimum entropy from the 3D one divided by the 3D entropy jump. We note that 1D minimum entropy, $s_{\rm min,1D}$, hardly depends on convection models, although this figure was produced using results with CM. The left and right panels are for the Eddington and VAL-C $T(\tau)$, respectively.}
  \label{fig:dsmin13d}
\end{figure*}

\begin{figure*}
  \begin{minipage}{0.49\textwidth}
    \centering
    \includegraphics[width=\hsize]{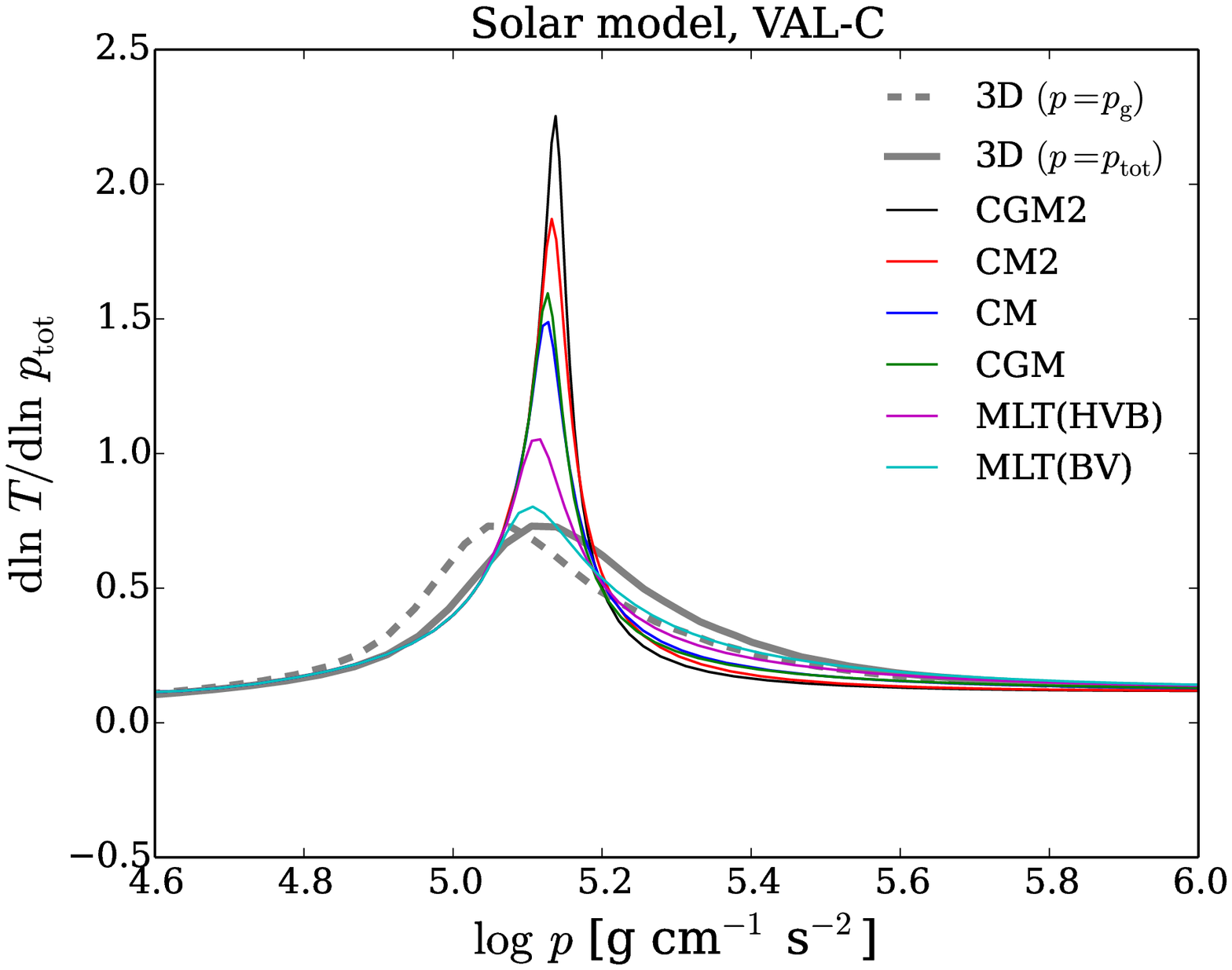}
  \end{minipage}
   \begin{minipage}{0.49\textwidth}
    \centering
    \includegraphics[width=\hsize]{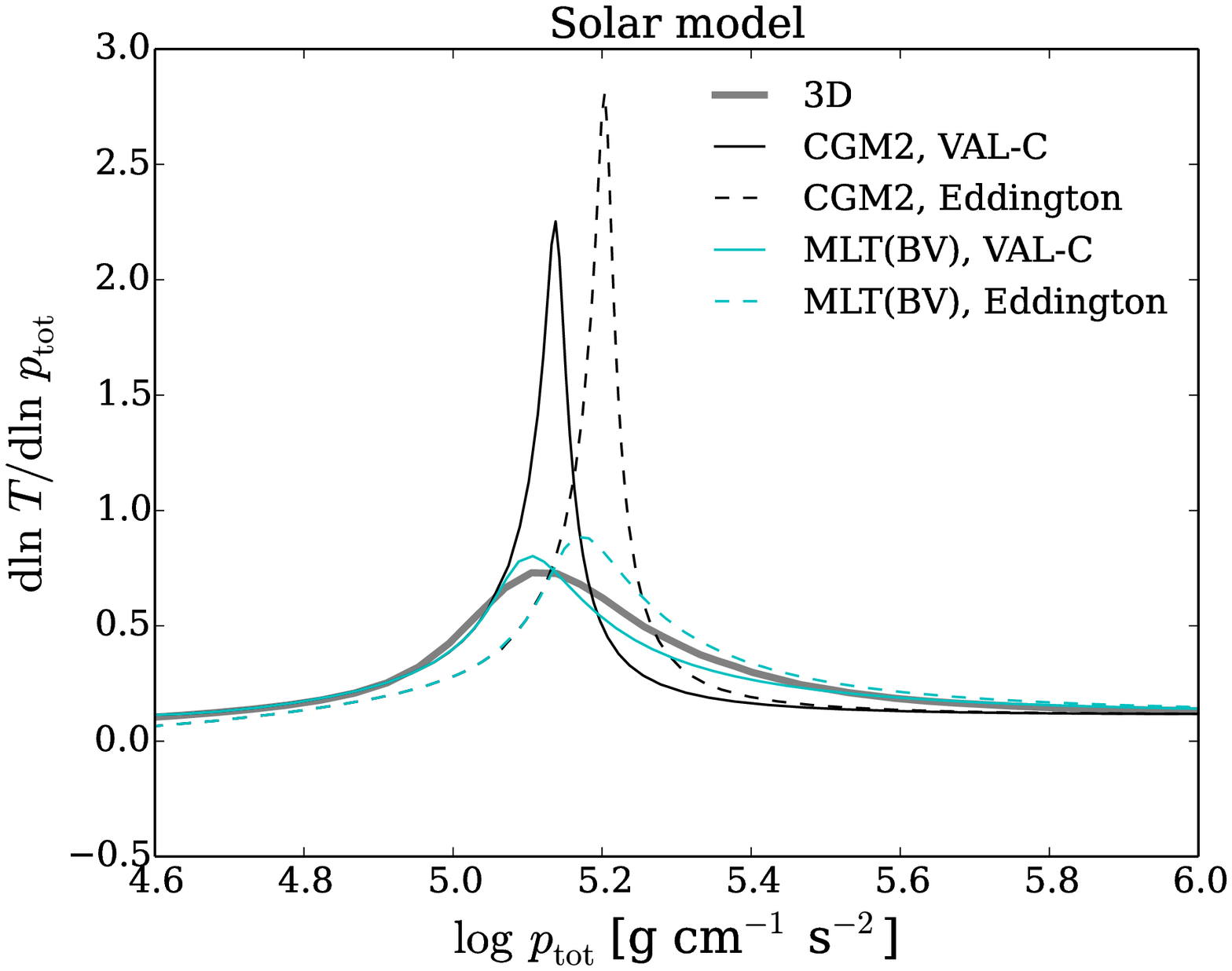}
  \end{minipage}
  \caption{Same as Fig. \ref{fig:diffconv}, but for comparison of temperature gradient profiles.}
  \label{fig:diffconv_nabla}
\end{figure*}

\begin{figure*}
  \begin{minipage}{0.49\textwidth}
    \centering
    \includegraphics[width=\hsize]{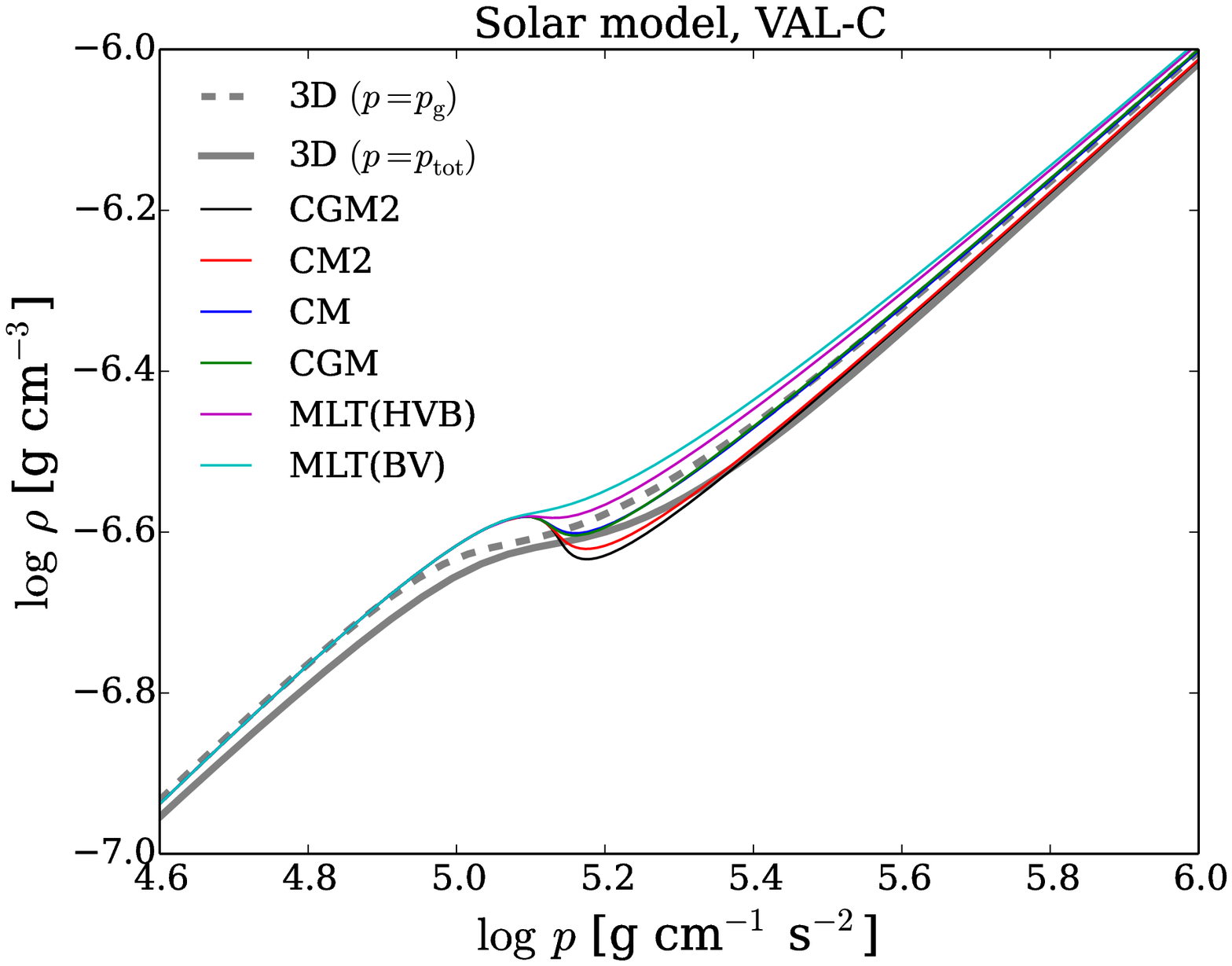}
  \end{minipage}
  \begin{minipage}{0.49\textwidth}
    \centering
    \includegraphics[width=\hsize]{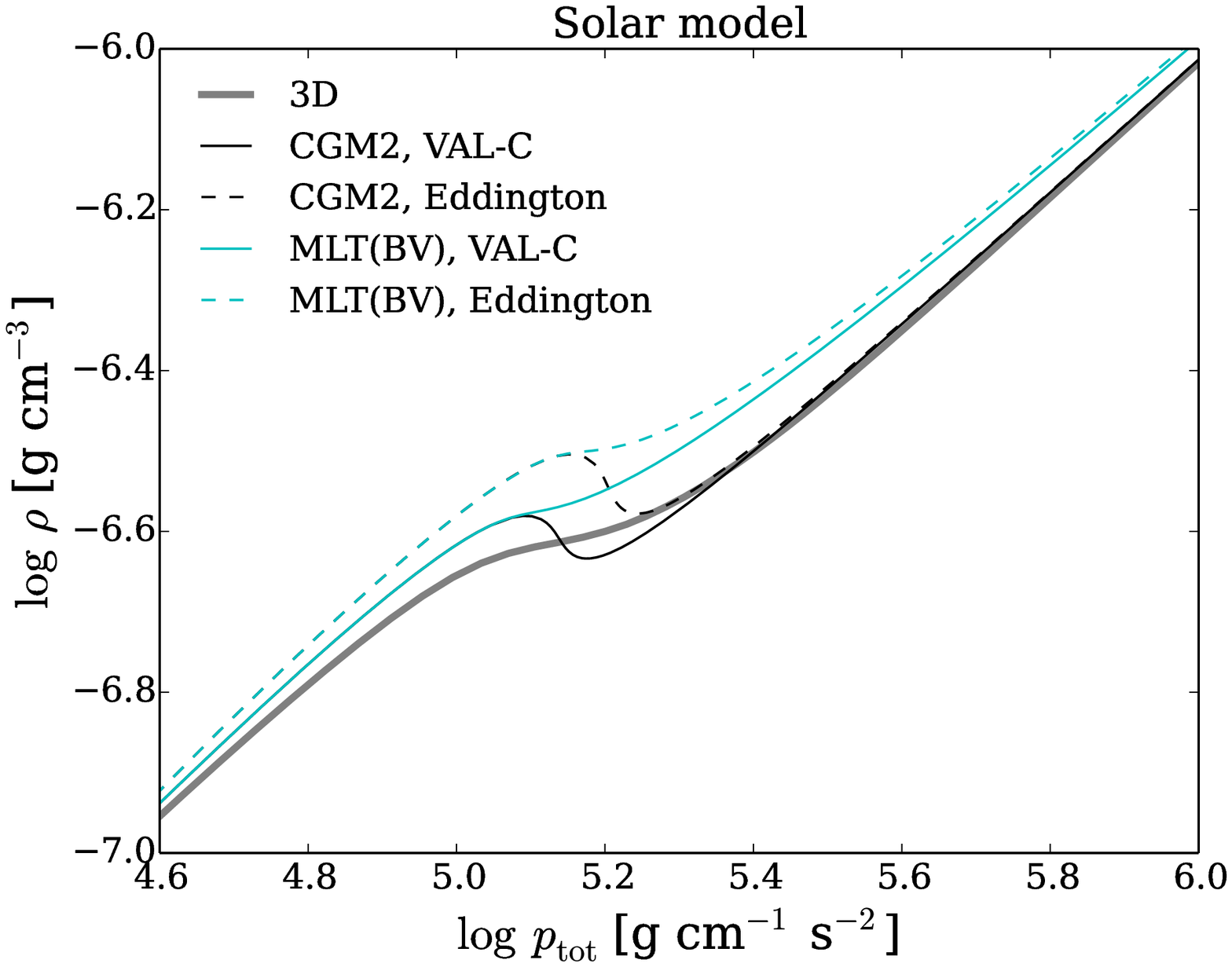}
  \end{minipage}
  \caption{Same as Fig. \ref{fig:diffconv}, but for comparison of density profiles.}
 \label{fig:diffconv_rho}
\end{figure*}

\begin{figure*}
  \begin{minipage}{0.49\textwidth}
    \centering
    \includegraphics[width=\hsize]{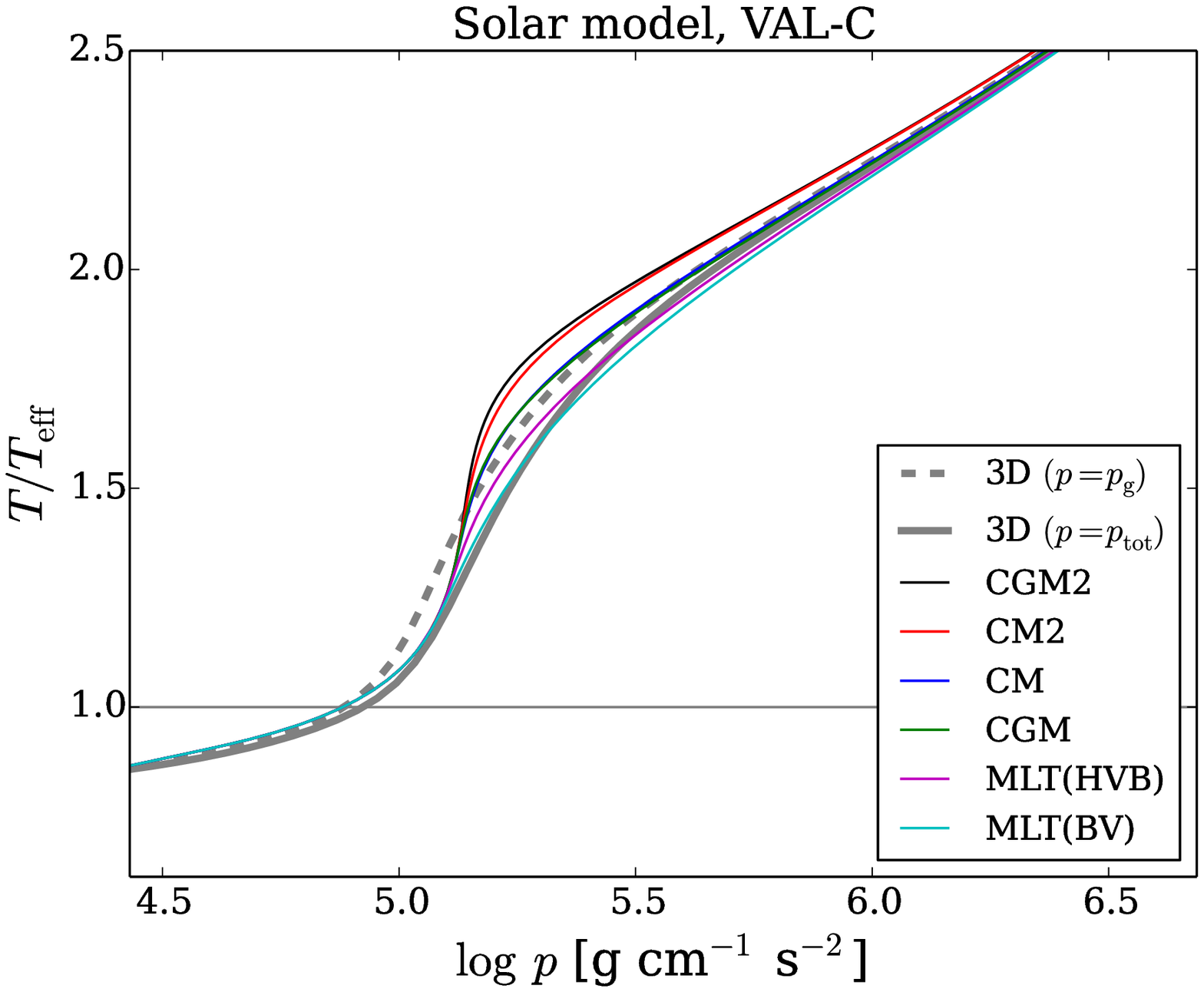}
  \end{minipage}
   \begin{minipage}{0.49\textwidth}
    \centering
    \includegraphics[width=\hsize]{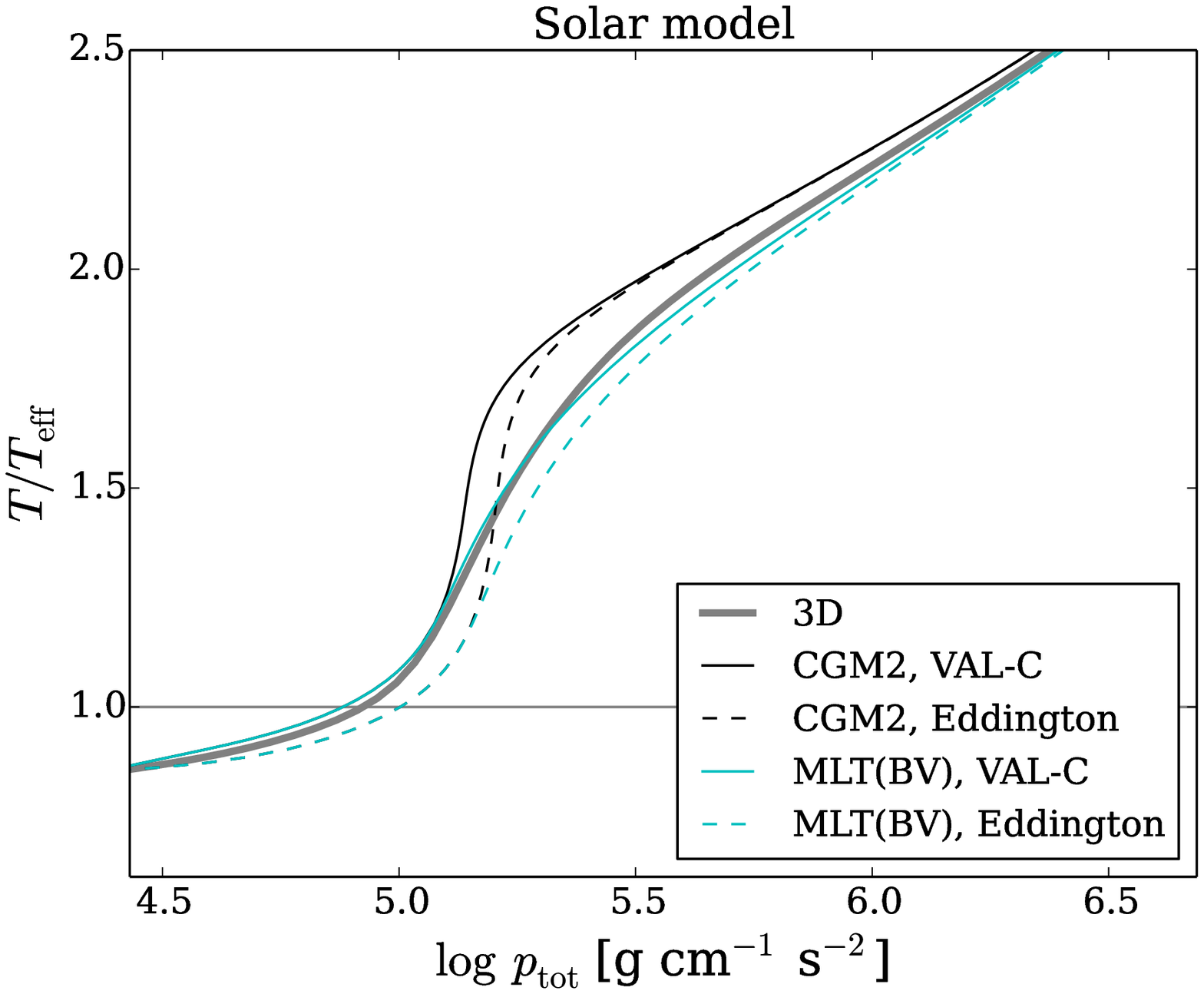}
  \end{minipage}
  \caption{Same as Fig. \ref{fig:diffconv}, but for comparison of temperature profiles normalized with the effective temperature.}
  \label{fig:diffconv_tteff}
\end{figure*}

\subsection{Calibrated 1D model profiles}
\label{sec:dev}

Figure \ref{fig:diffconv} compares the 1D entropy profiles with different convection models and $T(\tau)$ relations for the solar model. The 3D profile is also shown as functions of gas and total pressure. Since the latter case includes turbulent pressure, the profile shifts toward higher pressure for a fixed entropy value. The entropy profile in the radiative part of the atmosphere depends on $T(\tau)$ relations, while the convective inner part depends on convection models. In the following, we discuss each point.

\subsubsection{Equilibrium structure}
\label{sec:equil}

\paragraph{Dependence on $T(\tau)$-relations}
\ \\

\cite{Magic15} showed certain inverse correlations between $\alpha$ values and entropy jumps. While the top of an entropy jump is determined by the asymptotic entropy, which stems from a 3D model, the bottom of the jump, namely the photospheric-minimum entropy, depends on the $T(\tau)$ relation. In general, the VAL-C $T(\tau)$ gives better agreement of the photospheric-minimum entropy with the 3D profile than the Eddington $T(\tau)$. Figure \ref{fig:dsmin13d} shows that the deviation of the photospheric minimum from the 3D one is larger for the Eddington $T(\tau)$ than for the VAL-C $T(\tau)$. Particularly, the deviation is substantially large for low $T_{\rm eff}$ and high $g$ in case of the Eddington $T(\tau)$. For the Eddington $T(\tau)$, the photospheric minimum is generally smaller than for the VAL-C $T(\tau)$. Namely, the 1D entropy jump is larger, and hence a smaller $\alpha$ value is required to fit the 3D profile. We limit this work to models with $T_{\rm eff}\ga 4500$ K for $\log\,g\ga 2.5$. For lower $T_{\rm eff}$, the $T(\tau)$ relations adopted for solar-like stars may not be necessarily appropriate to represent a temperature profile, since H$_2$-molecule formation contributes to the atmosphere structure.

As shown in Sect. \ref{sec:alpha_fit}, the $\alpha$ value mostly correlates with the entropy jump. Therefore, the dependence of the $\alpha$ value on the $T(\tau)$ relation could be mainly explained by the difference in the photospheric-minimum entropy value itself. On the other hand, the optical depth at the entropy minimum varies slightly with $T_{\rm eff}$ and $g$. Such a variation has some effect on the entropy gradient and hence the $\alpha$ value. For both the $T(\tau)$ relations, roughly speaking, the entropy minimum shifts toward the larger optical depth side with decreasing $T_{\rm eff}$. For VAL-C, however, the depth at the entropy minimum more largely depends on $g$ than for the Eddington $T(\tau)$ relation. This may cause weaker correlations between the $\alpha$ value and the entropy jump. We discuss this point in Sect. \ref{sec:tausmin}.

\paragraph{Dependence on convection models}
\ \\

Below the photosphere, a 1D profile depends on the convection model. Figure \ref{fig:diffconv} shows that the FST models with $l=\alpha H_p$ (CM and CGM) have steeper entropy gradients than MLT. \cite{Canuto96} showed that the MLT overestimates the convective flux for low-efficiency convection, while underestimates for high-efficiency convection. Hence, it requires smaller temperature gradient in the superadiabatic layers, and higher gradient in the adiabatic bottom layers (Fig. \ref{fig:diffconv_nabla}). While the growth of small eddies is suppressed by radiation in the low-efficiency convection, different size eddies contribute to energy transport in the high-efficiency convection. The FST models reflect this tendency, but the single-eddy-size approximation that MLT adopts misses it.

CM2 and CGM2 have even steeper entropy gradients, and then the entropy approaches the asymptotic value more quickly if we compare at the same-pressure location. By the definition $l=r_{\rm top}-r+\alpha^*H_{p,{\rm top}}$, the mixing length becomes smaller as going upward. Indeed, the convective flux has high sensitivity to the mixing length \citep[$F_{\rm conv}\propto l^8$, see e.g.][]{Gough76} in the low efficiency limit. To compensate for this, the temperature gradient and hence the entropy jump should become steeper. Since the mixing length increases significantly with depth, the contrast of the entropy jump is extreme between the bottom and top of a convective zone.

Of the MLT models, the HVB version gives a slightly steeper entropy gradient than the BV version. As mentioned in Sect. \ref{sec:1D}, they differ in the assumption of the temperature distribution in a convection eddy. It makes the difference in the convective efficiency and hence in the required temperature and entropy gradients between the two versions.

While nonlocal effects are taken into account in the 3D models, the 1D models were constructed with a local treatment. This treatment forces the convective velocity to be zero at the Schwarzschild boundary. Near the top of a convective envelope, therefore, the convective velocity has a steep gradient. It leads to a pronounced density inversion as well as steep temperature gradient. Due to the smaller convective flux for low-efficiency convection, the FST models have steeper temperature gradient (Fig. \ref{fig:diffconv_nabla}) and hence larger density inversion than MLT (Fig. \ref{fig:diffconv_rho}). On the other hand, the 3D model has no density inversion.

The correspondence between 1D and 3D models significantly depends on physical quantities to be compared. For example, while the entropy profiles with CM and CGM match best the entropy profile of the 3D model as function of pressure (Fig. \ref{fig:diffconv}), the MLT models are closer to the 3D model in terms of the temperature (Fig. \ref{fig:diffconv_tteff}) and its gradient (Fig. \ref{fig:diffconv_nabla}). 
The averaged thermodynamic quantities of the 3D model are not related via an EOS, particularly in the convection zone where horizontal fluctuations are large. In contrast, in 1D models, they are related.
Therefore, it is hard to pursue the complete correspondence of all the quantities at the same time.
 
\subsubsection{Oscillation frequencies}

Although we discuss the difficulty in the comparison of the equilibrium structure among the 3D models and the calibrated 1D models in Sect. \ref{sec:equil}, comparing oscillation frequencies is expected to shed light on such an issue especially from the viewpoint of asteroseismology. To obtain appropriate frequencies, we combined the solar 1D and 3D envelope models with a 1D inner part model constructed by the CESTAM stellar evolution code \citep{Marques13}. The inner part model is similar to the one constructed by \cite{Sonoi15, Sonoi17}, but it was slightly improved so that the frequencies of the radial low-order modes in the resulting patched model match better with the observed ones given by \cite{Broomhall09}.

Concretely speaking, we searched for a model with the resulting-patched-model frequency closest to the observed one for the radial $n=6$ mode among CESTAM models with the solar mass, MLT(BV), the Eddington $T(\tau)$ and no turbulent pressure. We ranged the age from 4.96 to 5.05 Gyr and the mixing length parameter $\alpha(=l/H_p)$ from 1.648 to 1.652. We replaced the outer part of each model with the solar 3D model and computed adiabatic frequency of the radial $n=6$ mode using the ADIPLS code \citep{CD08}. We found that a model with an age 4.99 Gyr and $\alpha=$1.651 makes the frequency closest to the observed one. This model has a surface gravity acceleration $\log\,g=$4.439, which is close to that of the 3D model, $\log\,g=$4.438. Actually this model does not have the best connection between the 1D-inner and 3D-outer parts among the CESTAM models which we investigated. The temperature of these models is 30 to 50 K lower than that of the 3D model at the location with the pressure corresponding to the bottom of the 3D model. For the model with an age 4.99 Gyr and $\alpha=1.651$, the temperature is 41 K lower than the 3D model at this location. This difference is larger than but of the order of the convective fluctuation of the temperature in the 3D model, 23 K. The CESTAM models use the equation of states and opacity provided by OPAL2005 \citep{Rogers02, Iglesias96}, which are different from those of the 3D model. Such an inconsistency made it difficult to optimize all the properties at the same time.

\begin{figure}
  \centering
  \includegraphics[width=\hsize]{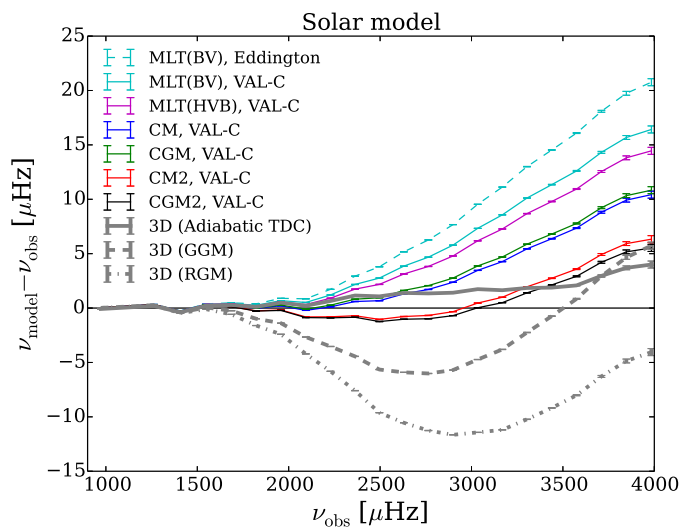}
  \caption{Difference between adiabatic radial-mode frequencies of the solar patched models with the 3D envelope and calibrated 1D envelope models and the observed ones given by \cite{Broomhall09} as function of the observed frequency. The error bars stem from the observed frequencies.}
  \label{fig:freq}
\end{figure}

We also constructed patched models with the calibrated 1D envelope models based on the inner-part model having 4.99 Gyr and $\alpha=1.651$. Using the ADIPLS code, we computed adiabatic radial-mode frequencies for the patched models in the whole range of the observed frequencies. For the patched model with the 3D model, which includes turbulent pressure, we computed frequencies with the reduced-gamma model (RGM) and gas-gamma model (GGM) approximations \citep{Rosenthal99}. While RGM neglects the perturbation of turbulent pressure, GGM assumes that the relative perturbation of turbulent pressure is equal to that of gas pressure. To consider the perturbation of turbulent pressure in a more physically-based way, the 3D-model frequencies were also computed using the MAD code \citep{Dupret02} with a time-dependent treatment of convection (TDC) for adiabatic pulsations introduced by \cite{Sonoi17} based on the formalism by \cite{Grigahcene05} and \cite{Dupret06}.

Figure \ref{fig:freq} shows the deviation of the patched model frequencies from the observed ones given by \cite{Broomhall09}. As for the TDC frequencies of the 3D model, the TDC formalism has a free parameter $\beta$ related to the closure problem. In this figure, we show TDC frequencies with $\beta=-0.44-0.34i$, which was found to result in frequencies close to the observed ones. Of the 1D-envelope-patched models, those for CM2 and CGM2 with VAL-C $T(\tau)$ are found to give the lowest frequencies, which are closest to the observed frequencies. CM and CGM follow them, and MLT results in the most deviating frequencies. We also see that the Eddington $T(\tau)$ gives higher frequencies than VAL-C. However, CM2 and CGM2 have much steeper entropy and temperature gradients and much stronger density inversions around the superadiabatic layers than the other convection models (Figs. \ref{fig:diffconv}, \ref{fig:diffconv_nabla} and \ref{fig:diffconv_rho}). Namely, their apparently appropriate frequencies are coming about by compensatory effects between deviating density and temperature profiles.

To solve the problem of the compensatory effects, the 1D models might need a larger acoustic cavity with moderate gradients of thermodynamical quantities by considering additional physics such as turbulent pressure and overshooting to have as low frequencies as the RGM frequencies of the 3D model without the perturbation of turbulent pressure. As for the effects of this perturbation, for now, we can see them by comparing the 3D frequencies of RGM, GGM and TDC. If we compare RGM and GGM, we see that the inclusion of this perturbation even in a crude way should increase the frequencies. The result with TDC shows that the more sophisticated way can bring the frequencies closer to the observed ones.

\subsection{Trends of calibrated $\alpha$ among different types of stars}
\label{sec:alpha_fit}

\subsubsection{Functional fits in the $T_{\rm eff}$--$\log\,g$ plane}
\label{sec:alpha_funcfit}

\begin{table*}
  \caption{Resulting coefficients $a_i$ of the fitting function $f$ (Eq. \ref{eq:fit}) to the calibrated $\alpha$ values, and the root-mean-square and maximal deviation of the fits from the the calibrated $\alpha$, rms$\Delta$ and max$\Delta$.}
  \label{tab:ai}
  \centering\footnotesize
  \begin{tabular}{llrrrrrrrcc}\hline\hline
    Conv.& $T(\tau)$ &  $a_0$   & $a_1$       & $a_2$     & $a_3$       & $a_4$       & $a_5$       & $a_6$ & $\Delta_{\rm rms}$ & $\Delta_{\rm max}$ \\
    \hline
    CM   & VAL-C     & 1.091537 & $-$0.073638 & 0.139087 &    0.029489 & $-$0.048683 & $-$0.008506 & $-$0.012394 & 0.013 & 0.029 \\
         & Eddington & 0.933675 & $-$0.131379 & 0.111231 &    0.032637 & $-$0.039236 &    0.043460 & $-$0.029604 & 0.011 & 0.026 \\
    \hline
    CGM  & VAL-C     & 0.821021 & $-$0.062016 & 0.110100 &    0.018825 & $-$0.029943 & $-$0.009310 & $-$0.006214 & 0.010 & 0.020 \\
         & Eddington & 0.703031 & $-$0.103263 & 0.087566 &    0.021193 & $-$0.023523 &    0.029304 & $-$0.019150 & 0.008 & 0.017 \\
    \hline
    CM2  & VAL-C     & 0.729778 & $-$0.075988 & 0.028712 &    0.051974 & $-$0.192901 &    0.144106 & $-$0.129416 & 0.037 & 0.081 \\
         &Eddington  & 0.539527 & $-$0.126751 & 0.002922 &    0.022234 & $-$0.124673 &    0.239318 & $-$0.123840 & 0.036 & 0.078 \\
    \hline
    CGM2 & VAL-C     & 0.359522 & $-$0.118199 & 0.061132 &    0.009828 & $-$0.122208 &    0.168391 & $-$0.124905 & 0.040 & 0.109 \\
         & Eddington & 0.225285 & $-$0.138345 & 0.030226 & $-$0.023650 & $-$0.065260 &    0.257248 & $-$0.117526 & 0.037 & 0.077 \\
    \hline
    MLT(BV) & VAL-C     & 2.057349 & $-$0.047880 & 0.107147 & $-$0.009432 &$-$0.011211 & $-$0.039115 &    0.012995 & 0.018 & 0.039 \\
            & Eddington & 1.790295 & $-$0.149542 & 0.069574 & $-$0.008292 &   0.013165 &    0.080333 & $-$0.033066 & 0.025 & 0.067 \\
    \hline
    MLT(HVB)& VAL-C     & 2.266096 & $-$0.086964 & 0.177333 & $-$0.001269 &$-$0.013706 & $-$0.039661 &    0.011885 & 0.020 & 0.045 \\
            & Eddington & 1.963503 & $-$0.197476 & 0.129092 &   0.007492 &    0.004862 &    0.075030 & $-$0.029763 & 0.024 & 0.054 \\
    \hline
  \end{tabular}
\end{table*}

We provide the calibrated $\alpha$ values for all the models in Table \ref{tab:alpha_m00}. In the CM2 and CGM2 cases, the calibrated $\alpha^*$ values for a few low-$T_{\rm eff}$ models are negative. These negative values are required to make entropy jumps high enough to obtain the asymptotic entropy matching. However, they lead to negative values of the mixing length $l$ near the top of the convective envelopes. In the layers with the negative $l$, we assumed purely radiative transfer.

\begin{figure}
  \centering
  \includegraphics[width=\hsize]{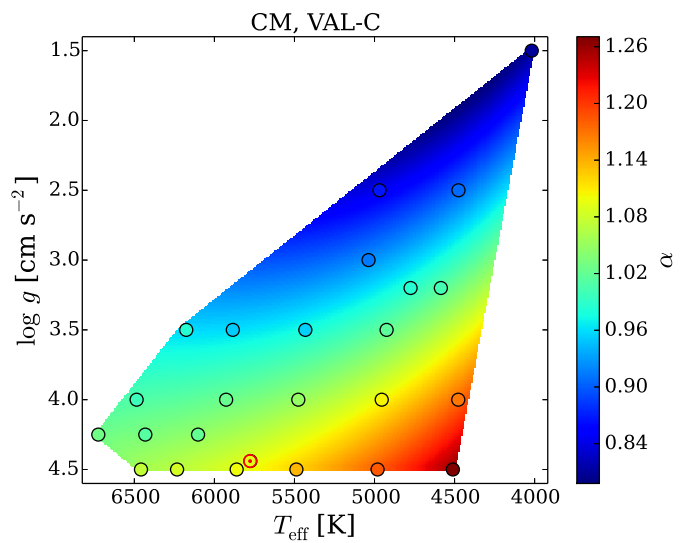}
  \caption{Calibrated $\alpha$ values (dots) and fitting function (Eq. \ref{eq:fit}, contour) for CM with the VAL-C $T(\tau)$ in the $T_{\rm eff}-\log\,g$ plane. The solar model is labeled as the solar mark.}
  \label{fig:alpha_cm_valc_m00}
\end{figure}

We performed functional fitting to the $\alpha$ values of all the models. Similarly to LFS and \cite{Magic15}, we fitted a function
\begin{eqnarray}
  f(x,y) = a_0 + (a_1+(a_3 + a_5x + a_6y)x + a_4y)x + a_2y
  \label{eq:fit}
\end{eqnarray} 
to the $\alpha$ values with a least-squares minimization method, where $x\equiv(T_{\rm eff}-5777)/1000$ and $y\equiv\log\,g-4.44$. Namely, $x$ and $y$ represent deviation from the solar effective temperature and surface gravity, respectively. As an example, the resulting fitting function for CM with the VAL-C $T(\tau)$ is shown as the color contour in Fig. \ref{fig:alpha_cm_valc_m00}. Resulting values of the coefficients $a_i$ for all the combinations of the convection models and $T(\tau)$ relations are given by Table \ref{tab:ai}. The last two columns show the deviation of the fitting function from the calibrated $\alpha$ values. The first one is the root-mean-square deviation $\Delta_{\rm rms}=\sqrt{\sum^N_i(f_i-\alpha_i)^2/N}$, where $N$ is the number of models, and the second one is the maximum of the absolute deviation, $\Delta_{\rm max}={\rm max}(|f_i-\alpha_i|)$. They are smaller for CM and CGM than for CM2, CGM2 and MLT. Namely, Eq. (\ref{eq:fit}) works better for CM and CGM.

\subsubsection{On correlation with entropy jump}
\label{sec:tausmin}

\begin{figure}
  \centering
  \includegraphics[width=\hsize]{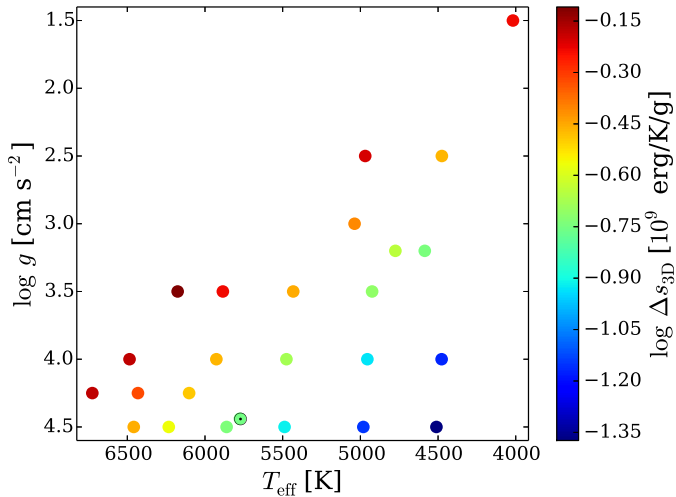}
  \caption{Entropy jump, which means difference between asymptotic entropy and photospheric minimum, in 3D models as a function of $T_{\rm eff}$ and $\log\,g$. The solar model is labeled as the solar mark.}
  \label{fig:lnsjump3d}
\end{figure}

\cite{Magic15} showed that MLT $\alpha$ values inversely correlate with the entropy jump. The $\alpha$ value increases with decreasing $T_{\rm eff}$ or increasing $g$, while the entropy jump decreases (Fig. \ref{fig:lnsjump3d}). The detailed comparison with \citeauthor{Magic15} will be discussed in Sect. \ref{sec:Magic15}.
  
\begin{figure}
  \centering
  \includegraphics[width=\hsize]{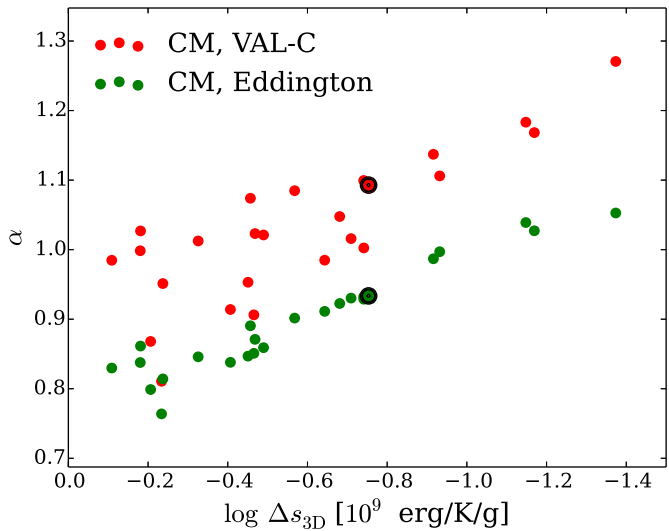}
  \caption{Calibrated $\alpha$ values for CM with different $T(\tau)$ relations as functions of $\log\,\Delta s_{\rm 3D}$. The solar marks indicate the solar model.}
  \label{fig:mlnsjump3d_vs_alpha_cm_m00}
\end{figure}

\begin{figure*}
  \begin{minipage}{0.49\textwidth}
    \centering
    \includegraphics[width=\hsize]{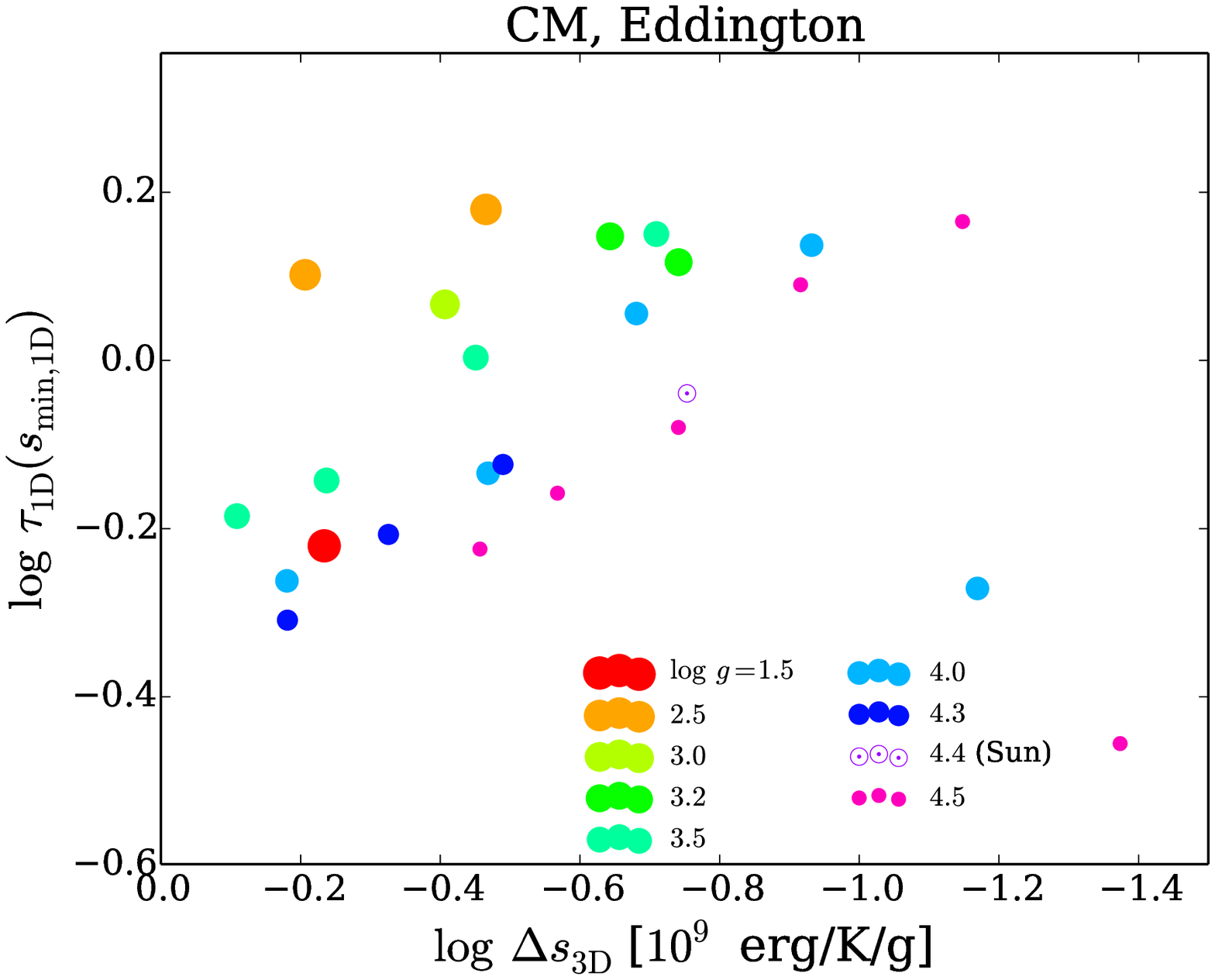}
  \end{minipage}
   \begin{minipage}{0.49\textwidth}
    \centering
    \includegraphics[width=\hsize]{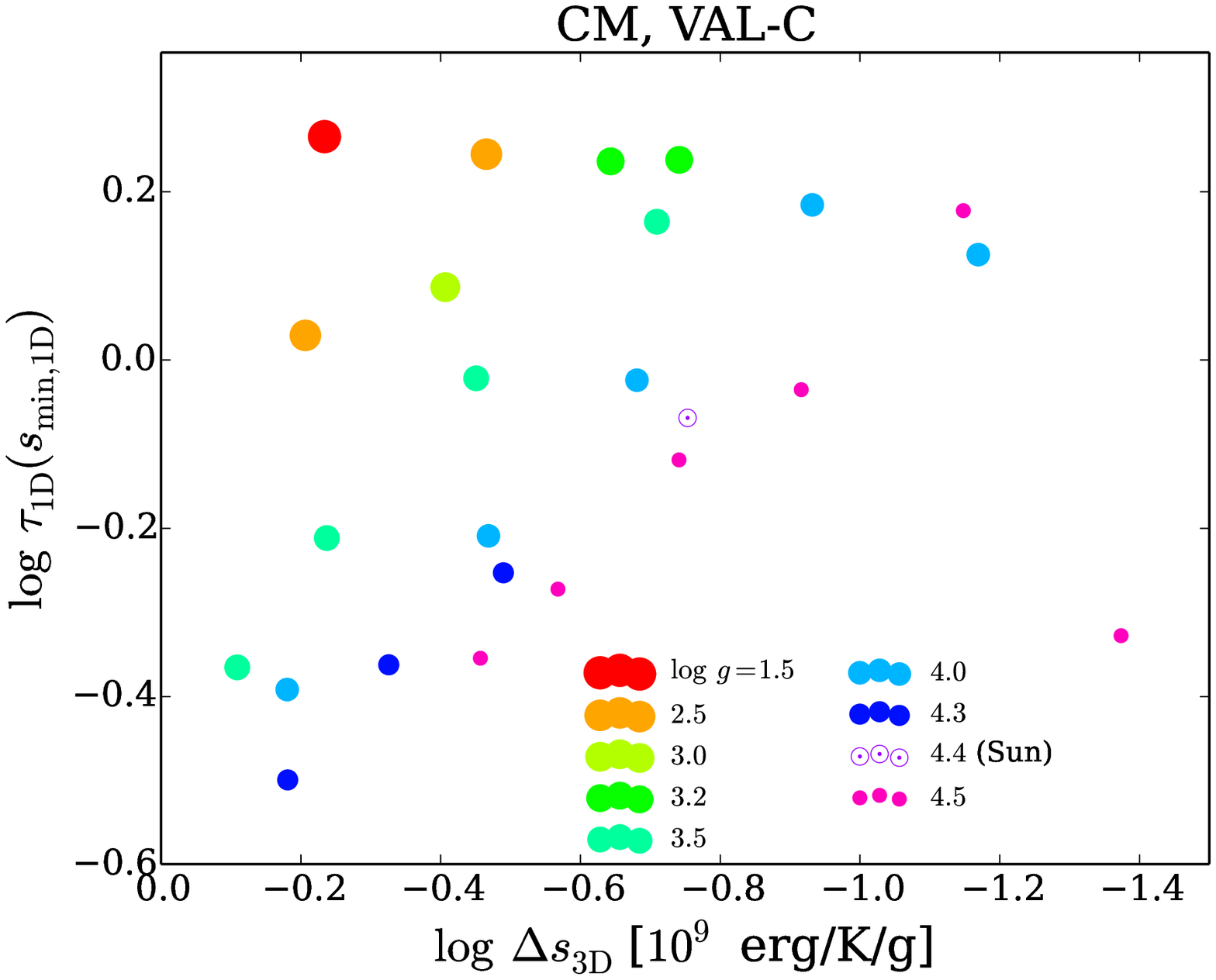}
  \end{minipage}
  \caption{Optical depth at the entropy minimum of the 1D model with the Eddington $T(\tau)$ (left) and VAL-C $T(\tau)$ (right). We note that the depth at the entropy minimum hardly depends on convection models similarly to the case of Fig. \ref{fig:dsmin13d}, although this figure was produced using results with CM.}
  \label{fig:logtausmin}
\end{figure*}

To check such a correlation, we plot the $\alpha$ values as function of the entropy jump in Fig. \ref{fig:mlnsjump3d_vs_alpha_cm_m00}, which shows the case of CM. Similarly to \citeauthor{Magic15}, the $\alpha$ value increases with decreasing entropy jump. This trend is generally found for all the convection models with $l=\alpha H_p$. However, the correlation is weaker for the VAL-C $T(\tau)$ than for the Eddington $T(\tau)$. Such a weaker correlation for VAL-C is also found for the other convection models as shown in Figs. \ref{fig:Magic15} and \ref{fig:Trampedach14}. It implies that additional factors would be required to explain the trend of $\alpha$ for the VAL-C $T(\tau)$. As a candidate for such a factor, we here consider the depth at the photospheric-minimum entropy in 1D models. Indeed, we can naturally expect that it could affect the entropy gradient and hence the $\alpha$ value. Figure \ref{fig:logtausmin} shows the optical depth at the entropy minimum as function of the entropy jump. For the VAL-C $T(\tau)$, the dependence on $g$ is stronger and hence the correlation with the entropy jump is weaker. This tendency is expected to be responsible for the weaker correlation between the entropy jump and $\alpha$ value compared to the Eddington $T(\tau)$ case (Fig. \ref{fig:mlnsjump3d_vs_alpha_cm_m00}).

\begin{figure}
  \centering
  \includegraphics[width=\hsize]{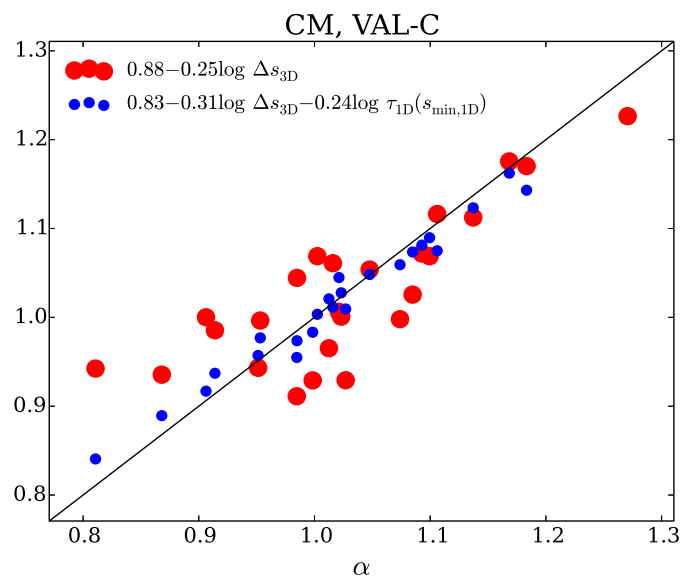}
  \caption{Correlation between $\alpha$ and the right hand side of Eq. (\ref{eq:abds}) (red) or (\ref{eq:abdsct}) (blue) for CM with the VAL-C $T(\tau)$.}
  \label{fig:alpha_vs_rhs}
\end{figure}

In order to see the dependence of $\alpha$ on the depth at the entropy minimum for CM with VAL-C, we performed linear functional fits as follows. By the least-squares method, we evaluated coefficients of functions
\begin{eqnarray}
  \alpha=a+b\log\,\Delta s_{\rm 3D}
  \label{eq:abds}
\end{eqnarray}
and
\begin{eqnarray}
  \alpha=a+b\log\,\Delta s_{\rm 3D}+c\log\,\tau_{\rm 1D}(s_{\rm min,1D}).
  \label{eq:abdsct}
\end{eqnarray}
We obtained $a=0.88$ and $b=-0.25$ for Eq. (\ref{eq:abds}), and $a=0.83$, $b=-0.31$ and $c=-0.24$ for Eq. (\ref{eq:abdsct}). The Pearson coefficients for the correlation between both sides of Eq. (\ref{eq:abds}) and (\ref{eq:abdsct}) were found to be 0.82 and 0.98, respectively. Namely, the inclusion of the term of $\log\,\tau_{\rm 1D}(s_{\rm min,1D})$ improves the correlation (see also Fig. \ref{fig:alpha_vs_rhs}). It implies that the weak correlation between the $\alpha$ value and the entropy jump for VAL-C can be caused by the fact that the depth at the entropy minimum has certainly different dependence on $T_{\rm eff}$ and $g$ from the entropy jump.

\subsubsection{Comparison with \cite{Magic15}'s results}
\label{sec:Magic15}

\begin{figure}
  \centering
  \includegraphics[width=\hsize]{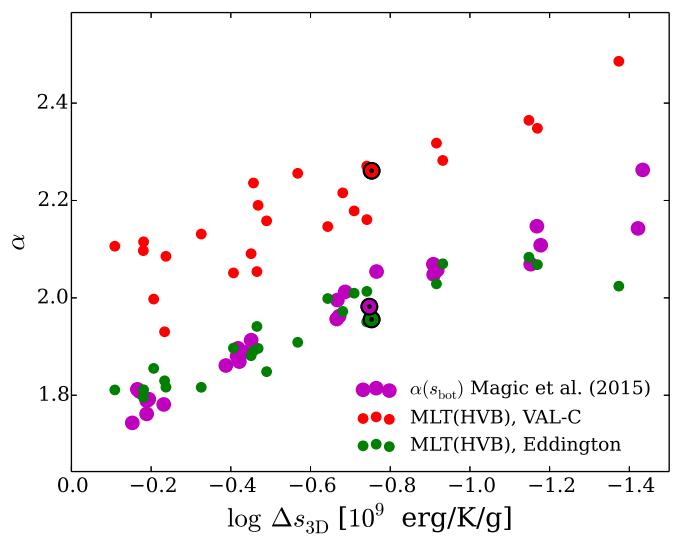}
  \caption{Comparison of our calibrated $\alpha$ values with those found by \cite{Magic15}. Similarly to \citeauthor{Magic15}, the HVB version of MLT is chosen to plot our $\alpha$ values. In \citeauthor{Magic15}, the radiative transfer is solved instead of using a fixed $T(\tau)$ relation for the 1D atmosphere. Although \citeauthor{Magic15} performed the $\alpha$ calibration in two ways, we choose to compare with the $\alpha$ values obtained by the asymptotic entropy matching, which are indicated as $\alpha(s_{\rm bot})$ in their paper (magenta dots). The solar marks indicate the solar model.}
  \label{fig:Magic15}
\end{figure}

Figure \ref{fig:Magic15} shows the $\alpha$ values of \cite{Magic15} as a function of $\log\,\Delta s_{\rm 3D}$. Similarly to the present work, their $\alpha$ values are calibrated by the fitting to the 3D asymptotic entropy, which are referred to $\alpha(s_{\rm bot})$ in their paper, but they also tried another method, fitting to the 3D entropy jump. The $\alpha$ values calibrated by the latter method are referred to $\alpha(\Delta s)$. Like our models, their 1D models do not include turbulent pressure. 

In Fig. \ref{fig:Magic15}, for comparison, we also plot our $\alpha$ values for the HVB version of MLT, which was adopted by \cite{Magic15}. While the present work adopts fixed $T(\tau)$ relations, \citeauthor{Magic15} used a 1D atmosphere code \citep{Magic13} which solves the radiative transfer equations instead of using a fixed $T(\tau)$ relation, as mentioned in Sect. \ref{sec:1D}. They demonstrated that $\alpha$ values obtained by the 1D atmosphere code are generally similar to those obtained by another 1D {\it envelope} code with the Eddington $T(\tau)$ \citep{CD08}. The same EOS and opacity package as their 3D models were implemented into both their atmosphere and envelope codes. While the $\alpha$ value by \citeauthor{Magic15}'s 1D atmosphere code solving radiative transfer monotonically increases with decreasing entropy jump, our $\alpha$ for the Eddington $T(\tau)$ drops in the lower limit of the entropy jump. This is because, in our 1D models with the Eddington $T(\tau)$, the photospheric-minimum entropy is much lower than the 3D one, the entropy jump is higher, and hence the $\alpha$ value becomes lower. Except in the lower limit of the entropy jump, their $\alpha$ values obtained with the 1D atmosphere code solving the radiative transfer are found to be close to those for our models with the Eddington $T(\tau)$. It is consistent with \citeauthor{Magic15}'s comparison with their another envelope code with the Eddington $T(\tau)$, which is mentioned above. \citeauthor{Magic15} reported that the $\alpha(\Delta s)$ values are systematically higher than the $\alpha(s_{\rm bot})$ since the photospheric-minimum entropy values of their 1D models are lower than those of their 3D models. This situation is similar to our 1D models with the Eddington $T(\tau)$. It implies that every sophisticated 1D atmosphere model does not necessarily give a superior correspondence of an entropy profile to a 3D model.

\subsubsection{Comparison with \cite{Trampedach14b}'s results}
\label{sec:Trampedach14}

\begin{figure}
  \centering
  \includegraphics[width=\hsize]{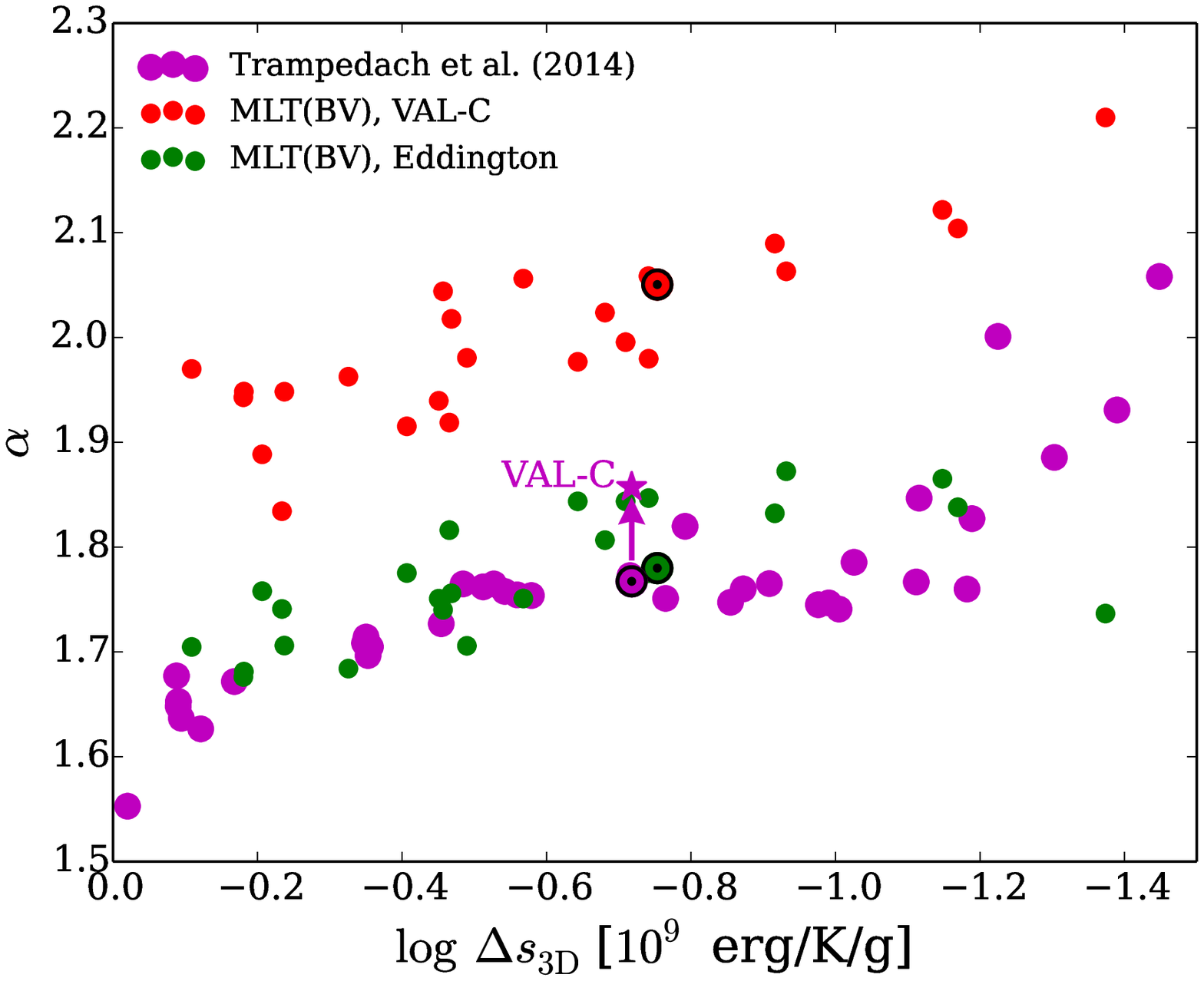}
  \caption{Comparison of the calibrated $\alpha$ values with \cite{Trampedach14b}. Similarly to \citeauthor{Trampedach14b}, the BV version of MLT is chosen to plot our $\alpha$ values. In \citeauthor{Trampedach14b}, the $T(\tau)$ relation based on their 3D models \citep{Trampedach14a} is adopted for constructing the atmosphere part of their 1D models. Unlike the present work, their 1D models include turbulent pressure and the $\alpha$ calibration was performed by matching temperature in the adiabatic bottom part of a convective envelope. Their $\alpha$ values obtained by the above procedure are presented as the magenta dots. They also tested with the VAL-C $T(\tau)$ relation, and reported that the $\alpha$ value becomes 0.09 larger for the solar model than the case of the $T(\tau)$ relation based on the 3D models (the magenta star). The solar marks indicate the solar model.}
  \label{fig:Trampedach14}
\end{figure}

Figure \ref{fig:Trampedach14} compares calibrated $\alpha$ values obtained by \cite{Trampedach14b}, who performed $\alpha$ calibration using the BV version of MLT for constructing 1D models, with our $\alpha$ values for the same convection model. Unlike \cite{Magic15} and the present work, they included turbulent pressure in the 1D models. While they performed the matching with the 3D models at the adiabatic bottom part of a convective envelope similarly to us, they matched temperature instead of entropy, and also turbulent pressure. For the 1D atmosphere, they used a $T(\tau)$ relation based on their 3D models \citep{Trampedach14a}. Also in this case, our values for the Eddington $T(\tau)$ are closer to their values than those for the VAL-C $T(\tau)$. However, the inclusion of turbulent pressure in their 1D models and the difference in the matching procedure make the comparison complicated.

\citeauthor{Trampedach14b} calibrated the $\alpha$ values also using the VAL-C $T(\tau)$. They found that VAL-C generally results in larger $\alpha$ values than the $T(\tau)$ relation based on the 3D models. They reported that VAL-C makes the $\alpha$ value 0.09 larger for the solar model (the magenta star in Fig. \ref{fig:Trampedach14}). Even in case of VAL-C, their $\alpha$ value is certainly smaller than ours. This implies that the different matching procedure mentioned above could be responsible for the difference between \citeauthor{Trampedach14b}'s and our $\alpha$ values.

\subsubsection{Comparison with LFS's results}
\label{sec:LFS}

The calibration with 2D models by LFS also adopted the BV version of MLT. However they warned their underestimate of $\alpha$ values due to a systematic effect of the 2D approximation. They suggested that the values should be multiplied by a factor slightly larger than unity to correspond to 3D cases. Indeed, \citeauthor{Trampedach14b} confirmed that their results agree with LFS's in the solar vicinity after multiplying by 1.11.

\begin{figure}
  \centering
  \includegraphics[width=\hsize]{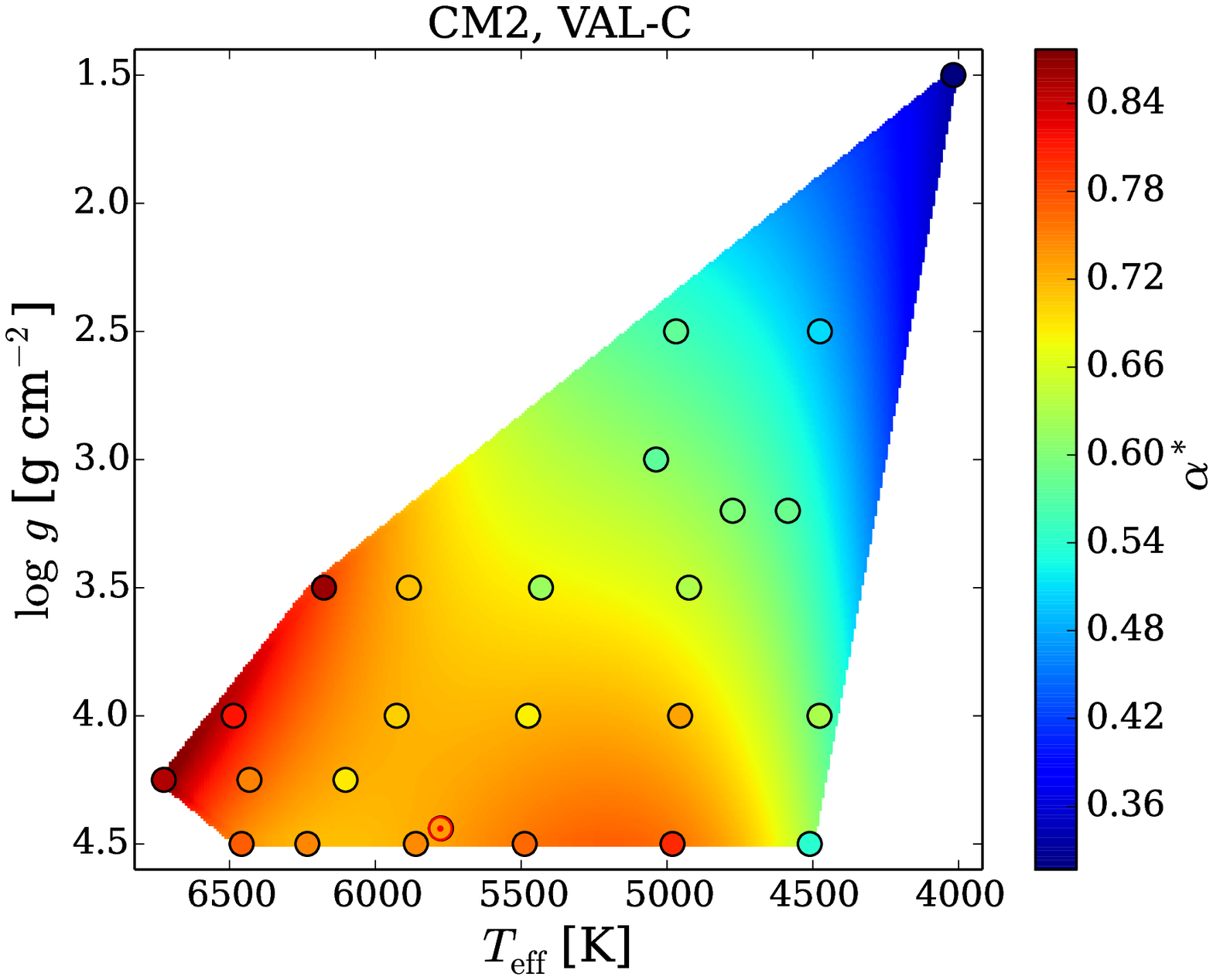}
  \caption{Same as Fig. \ref{fig:alpha_cm_valc_m00}, but for CM2 with the VAL-C $T(\tau)$}
  \label{fig:alpha_cm2_valc_m00}
\end{figure}

In addition to MLT, LFS performed $\alpha$ calibration for CM with definition of the mixing length, $l=r_{\rm top}-r+\alpha^* H_{p,{\rm top}}$, namely CM2. Similarly to their result, our calibrated $\alpha^*$ of CM2 and CGM2 increases with increasing $T_{\rm eff}$ or $g$ as shown in Fig. \ref{fig:alpha_cm2_valc_m00}. As mentioned in Sect. \ref{sec:asol}, the meaning of $\alpha^*$ is different from the case of $l=\alpha H_p$, and may be interpreted as a kind of overshooting above a convective envelope. LFS and the present work show that higher values are required at the high $T_{\rm eff}$ side.

\section{Discussion}
\label{sec:discussion}

\subsection{Stellar evolution with varying $\alpha$ based on the 3D calibration}

\begin{figure}
  \centering
  \includegraphics[width=\hsize]{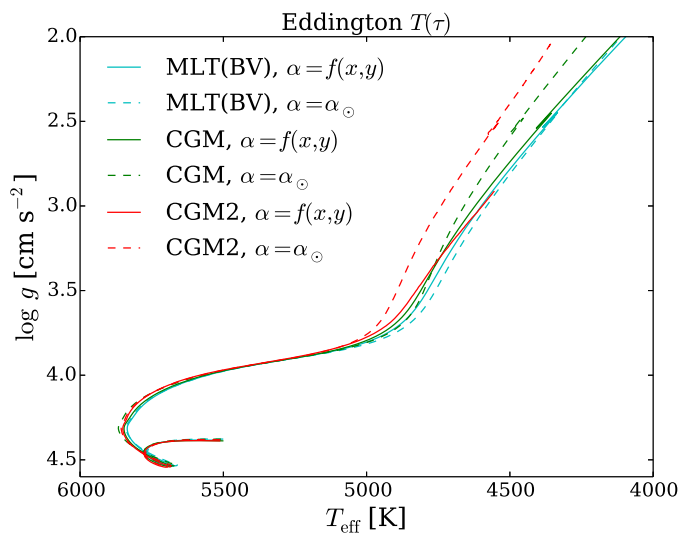}
  \caption{$1M_\odot$ evolutionary tracks obtained with the $\alpha$ values calibrated with the Eddington $T(\tau)$ relation. The solid lines are tracks with varying $\alpha$ which follows Eq. (\ref{eq:fit}). The dashed lines are ones with $\alpha$ fixed to the solar values (Table \ref{tab:sol}). In case of CGM2, the calibrated $\alpha^*$ is negative for low $T_{\rm eff}$ and $g$ as mentioned in Sect. \ref{sec:alpha_funcfit}. In this figure, we truncated the negative-$\alpha^*$ part of the evolutionary track with varying $\alpha^*$ for CGM2.}
  \label{fig:evol_edd}
\end{figure}

We tested evolutionary computation of a $1M_\odot$ star with varying $\alpha$ which follows Eq. (\ref{eq:fit}) using the code ATON 3.1. Figure \ref{fig:evol_edd} shows evolutionary tracks with the $\alpha$ values calibrated for MLT(BV), CGM and CGM2 with the Eddington $T(\tau)$. The dashed lines are tracks with $\alpha$ fixed to the solar value (Table \ref{tab:sol}), and the solid lines are those with varying $\alpha$ which follows Eq. (\ref{eq:fit}). The varying-$\alpha$ tracks for the different convection models and $T(\tau)$ relations should be identical to each other since they are based on the same grid of the 3D models. In practice, they commonly have $T_{\rm eff}\simeq 4100$ K at $\log\,g=2$, but they are not exactly identical due to the uncertainty on the interpolation of the calibrated $\alpha$ values in the sparse grid of the 3D models. As shown in the last two columns of Table \ref{tab:ai}, the deviation of the fitting function $f(x,y)$ from the calibrated $\alpha$ ranges from $\sim$0.01 to $\sim$0.1 depending on the convection models and $T(\tau)$ relations.

\begin{figure}
  \centering
  \includegraphics[width=\hsize]{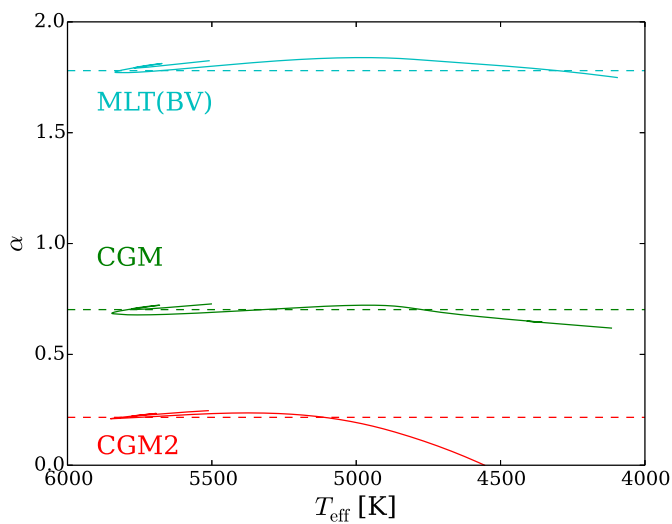}
  \caption{Variation of $\alpha$ on the evolutionary tracks shown in Fig. \ref{fig:evol_edd}. The horizontal dashed lines indicate the solar values (Table \ref{tab:sol}). Same as Fig. \ref{fig:evol_edd}, the negative-$\alpha^*$ part of the varying-$\alpha^*$ evolution for CGM2 is truncated.}
  \label{fig:avar}
\end{figure}

Figure \ref{fig:avar} shows the variation of the $\alpha$ values on the evolutionary tracks with varying $\alpha$. The difference in the tracks appears mainly in the low $T_{\rm eff}$ side due to the deviation of the calibrated $\alpha$ values from the solar value.

For MLT, the difference between the tracks with the solar $\alpha$ and the varying $\alpha$ is much smaller than for CGM and CGM2. Some recent papers have also reported such a small difference for MLT. Using the 3D-calibrated $\alpha$ for MLT(BV) provided by \cite{Trampedach14b}, \cite{Salaris15} and \cite{Mosumgaard18} computed evolution with varying $\alpha$. For $1M_\odot$ star evolution, \cite{Salaris15} found that the RGB track hardly differs between the cases of the solar $\alpha$ and varying $\alpha$ when the 3D-calibrated $T(\tau)$ relation \citep{Trampedach13} was used for both the cases. Rather they found larger shifts of the RGB track by up to 100 K when comparing the varying-$\alpha$ track with the 3D $T(\tau)$ relation and tracks with some fixed $T(\tau)$ relations and $\alpha$ value fixed to 1.76, the solar value provided by \cite{Trampedach14b}. The result that such small differences between the solar-$\alpha$ and varying-$\alpha$ tracks have been commonly found by \cite{Salaris15} and the present work can be explained by the fact that our 3D-calibrated $\alpha$ values for MLT(BV) with the Eddington $T(\tau)$ are close to those by \cite{Trampedach14b}, which is shown in Sect. \ref{sec:Trampedach14}.

\cite{Mosumgaard18} also found that a $T_{\rm eff}$ shift from a solar-$\alpha$ track with the Eddington $T(\tau)$ to a varying-$\alpha$ track with the 3D $T(\tau)$ relation \citep{Trampedach13} is not substantial, $\sim +30$ K at $\log\,g=3.2$ for $1M_\odot$. We find it similar to the $T_{\rm eff}$ shift from our solar-$\alpha$ to varying-$\alpha$ track for MLT, which is $\sim +24$ K at the same $\log\,g$. This similarity also can be explained by the closeness between \cite{Trampedach14b}'s $\alpha$ values and our 3D-calibrated $\alpha$ values for MLT(BV) with the Eddington $T(\tau)$.

For CGM and CGM2, the $T_{\rm eff}$ shift from the solar-$\alpha$ to varying-$\alpha$ track in the upper part of the RGB tracks is much larger than for MLT. For CGM, the shift is found to be $\simeq -40$ K at $\log\,g=3$ and $\simeq -120$ K at $\log\,g=2$. For CGM2, the varying $\alpha$ is reduced to zero at $\log\,g\simeq 3$, and then the track is truncated there in Fig. \ref{fig:evol_edd}. The $T_{\rm eff}$ shift is found to be $\simeq -140$ K at $\log\,g=3$. Such larger $T_{\rm eff}$ shifts can be explained by the larger deviation of the varying $\alpha$ value from the solar value shown in Fig. \ref{fig:avar}.

\subsection{Impacts of treatments of transition between optically thin and thick layers on $\alpha$ values}
\label{sec:nabrad}

In our 1D envelope code, the radiative temperature gradient, $\nabla_{\rm rad}$, is evaluated by Eq. (\ref{eq:nabla_rad}) in both optically thin and thick layers. Eq. (\ref{eq:nabla_rad}) corresponds to the differentiation of a $T(\tau)$ relation (Eq. \ref{eq:Ttau}). In the optically thick layers, it is equivalent with the diffusion approximation since ${\rm d}q(\tau)/{\rm d}\tau\rightarrow 0$ ($\tau\rightarrow\infty$). Around the photosphere, both of them contribute to the same degree. Namely, Eq. (\ref{eq:nabla_rad}) is important to give smooth transition between these two conditions. In addition, convection is considered consistently from the atmosphere part to the deep interior of a convective envelope when $\nabla_{\rm rad}$ exceeds the adiabatic temperature gradient.

In some evolution codes, however, a $T(\tau)$ relation is used assuming purely radiative transfer in the atmosphere. In the interior, $\nabla_{\rm rad}$ is evaluated exclusively with the diffusion approximation by neglecting the term ${\rm d}q(\tau)/{\rm d}\tau$ in Eq. (\ref{eq:nabla_rad}) while a convection model is used to obtain the actual temperature gradient when $\nabla_{\rm rad}$ is larger than the adiabatic gradient. Here, we checked the influence of such a treatment on the entropy profiles and the $\alpha$ values. We define the optical depth of the boundary between the atmosphere and inner parts as $\tau_\star$. We tested $\tau_\star=$0.1, 2/3 and 10 using CM with the VAL-C $T(\tau)$ for the solar model.

\begin{figure}
  \centering
  \includegraphics[width=\hsize]{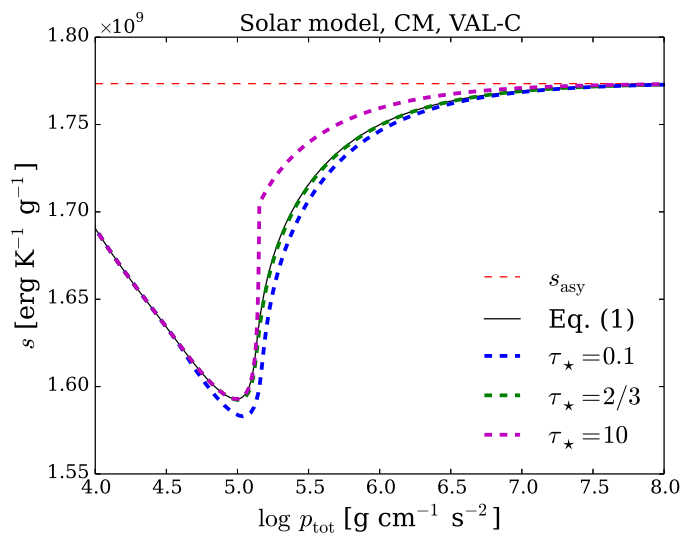}
  \caption{Same as Fig. \ref{fig:diffalpha}, but for comparison of different ways to evaluate $\nabla_{\rm rad}$. The profiles are 1D models calibrated using CM with the VAL-C $T(\tau)$. The black solid line was obtained by the method adopted to the 3D $\alpha$ calibration, using Eq. (\ref{eq:nabla_rad}) and considering convection both in the optically thin and thick layers. The dashed lines were obtained by a method adopted by some evolution codes (See the text).}
  \label{fig:s_difftaustar}
\end{figure}

Figure \ref{fig:s_difftaustar} compares a profile given by Eq. (\ref{eq:nabla_rad}) (the black solid line) and ones obtained with the method mentioned above (the dashed lines). $\tau_\star=2/3$ gives a profile close to the one by Eq. (\ref{eq:nabla_rad}). In case of $\tau_\star=0.1$, the photospheric-minimum entropy shifts toward the interior. Namely, the top of the convection zone shifts into a deeper layer. This is caused by the underestimate of $\nabla_{\rm rad}$ in the layers with $\tau>0.1$. In case of $\tau_\star=10$, the photospheric minimum is naturally close to the one by Eq. (\ref{eq:nabla_rad}). However, the convection model is not used while $\tau<10$ even if $\nabla_{\rm rad}>\nabla_{\rm ad}$. Therefore, the temperature gradient is overestimated, and hence the entropy increases abruptly just before $\tau$ reaches $\tau_\star$.

\begin{table}
  \caption{Comparison of $\alpha$ values obtained in different ways to evaluate $\nabla_{\rm rad}$.}
  \label{tab:difftaurad}
  \begin{center}
    \begin{tabular}{ccccc}\hline\hline
      & Eq.(\ref{eq:nabla_rad}) & $\tau_\star=0.1$ & $\tau_\star=2/3$ & $\tau_\star=10$ \\
      \hline
      $\alpha$ & 1.093                   & 0.985           & 1.079           & 1.600 \\
      \hline
    \end{tabular}
  \end{center}
\end{table}

Table \ref{tab:difftaurad} compares the calibrated $\alpha$. The $\alpha$ value for $\tau_\star=2/3$ is closest to the one obtained by Eq. (\ref{eq:nabla_rad}). For $\tau_\star=0.1$, since the photospheric minimum is smaller, a larger entropy jump is required, and hence the $\alpha$ value should be smaller. However the deviation from the entropy profile by Eq. (\ref{eq:nabla_rad}) is not so large. For $\tau_\star=10$, the deviation is much larger. Since the assumption of the purely radiative transfer even in the deeper layers gives so steep entropy gradient, the convective part ($\tau>\tau_\star$) does not need to make entropy jump so much. The $\alpha$ value is therefore much larger.

Although we have been discussing a case where the purely radiative transfer is assumed in the atmosphere, some evolution codes have the option to use a model atmosphere in which convection is taken into account. In such a case, the boundary between the atmosphere and inner parts should be set at a deeper layer, say $\tau_\star\sim 10-100$ \citep[e.g.][]{Morel94}. However, atmosphere grids are usually built with a fixed $\alpha$ value, which is in general calibrated with the Sun. If an $\alpha$ value or convection model of the atmosphere grids is different from the one used in the inner part, the impact of such an inconsistency in the treatment of convection between the atmosphere and inner parts must be assessed \citep{Montalban01,Montalban04}.

\section{Conclusion}
\label{sec:conclusion}

We calibrated the mixing-length parameters for convection models involved in 1D stellar evolution models with 3D models of solar-like stars produced by the CO$^5$BOLD code. We analyzed the classical mixing-length theory (MLT) proposed by \cite{BV58}, another version by \cite{Henyey65} and the full spectrum turbulence (FST) models proposed by \citeauthor{CM91} (\citeyear{CM91}, CM) and Canuto, Goldman \& Mazzitelli (\citeyear{CGM96}, CGM). For CM and CGM, we carried out additional calibrations using a mixing length of the form $l=r_{\rm top}-r+\alpha^*H_{p,{\rm top}}$ in addition to $l=\alpha H_p$.

In each case, we consider two outer conditions: the Eddington grey $T(\tau)$ relation and the one with the solar-calibrated Hopf-like function based on Model C of \citeauthor{Vernazza81} (\citeyear{Vernazza81}, VAL-C). The VAL-C $T(\tau)$ is found to give better correspondence to the 3D photospheric-minimum entropy than the Eddington $T(\tau)$ independently of convection models, since the energy transfer is totally radiative above the atmosphere except the vicinity of the photosphere.

On the other hand, the structure below the  photosphere depends on the convection model. CM and CGM make steeper entropy gradients than MLT. The definition of a mixing length $l=r_{\rm top}-r+\alpha^*H_{p,{\rm top}}$ gives a much steeper gradient than $l=\alpha H_p$, which agrees with a result shown by the CGM paper.

We find that not a single convection model gives the best correspondence to a 3D model in a convection zone. For example, while CM and CGM give the best correspondence to the 3D entropy profile, they are not necessarily the best and MLT comparably gives better correspondence to the 3D temperature profile. Unlike 1D models, averaged 3D quantities are not necessarily related via an EOS due to the convective fluctuation. Therefore, it is hard to pursue the complete correspondence of all the quantities at the same time. We compared radial-mode frequencies of the solar models to which the 3D envelope and the calibrated 1D envelope models are patched. We found that CM and CGM with $l=r_{\rm top}-r+\alpha^*H_{p,{\rm top}}$ give the frequencies closest to the observed ones. In the 1D models, however, the local treatment forces the convective velocity to have extremely steep gradient near the top of the convective envelope. It leads to a density inversion in the 1D models while the 3D models do not have it. The apparently appropriate frequencies of CM and CGM with $l=r_{\rm top}-r+\alpha^*H_{p,{\rm top}}$ are caused by the compensatory effects between the deviating density and temperature profiles. It implies that the inclusion of additional physics such as turbulent pressure and overshooting might be required.

We tested evolutionary computation with varying $\alpha$ based on our 3D-calibrated values. For MLT, the change from the solar $\alpha$ to the varying $\alpha$ hardly shifts the evolutionary track, which is consistent with the other papers. For CGM, on the other hand, the shift is found to be much larger particularly in the red giant stage. This implies that stellar evolutionary computations should not be computed fixing the mixing-length parameter to the solar $\alpha$ value but should rather adopt 3D-calibrated $\alpha$ values particularly for CGM. We provide tables of the calibrated $\alpha$ values (Table \ref{tab:alpha_m00}) and the coefficients of the fitting functions in the $T_{\rm eff}-\log\,g$ plane (Table \ref{tab:ai}). They should be implemented into evolutionary computation without turbulent pressure since our calibration is performed using 1D models without it. In a future work, we will perform $\alpha$ calibrations using 1D models with turbulent pressure. Although we calibrated $\alpha$ values by the asymptotic entropy matching in the present work, a calibration with another matching procedure adopted by \cite{Trampedach14b} would be preferable in order to understand the difference between their and our $\alpha$ values. We limited the present calibration to the solar metallicity. The other metallicity cases should be analyzed in a future work. \cite{Magic15} investigated the different metallicity cases. They found that the $\alpha$ value decreases by $\sim$ 0.1 to 0.2 while the metallicity increases from [Fe/H]$=-3.0$ to $+0.5$. The dependence on EOS and opacity tables is also worth analyzing.

Although the VAL-C $T(\tau)$ is found to give good correspondence to the photospheric-minimum entropy of 3D models, we stress that we limited this work to $T_{\rm eff}\ga 4500$ K for $\log\,g\ga 2.5$. For lower $T_{\rm eff}$, H$_2$ formation is important for a temperature profile in atmosphere, and a more appropriate $T(\tau)$ relation should be required. As possible candidates, $T(\tau)$ relations based on RHD models such as the ones proposed by \cite{Trampedach13} would be more suitable although even such $T(\tau)$ relations do not necessarily reproduce atmosphere structures exactly identical to RHD models \citep{Ludwig08}.

\begin{acknowledgements}
  We acknowledge the anonymous referee for helpful comments to improve this article.
  We are grateful to Regner Trampedach for fruitful discussions.
  T.S. has been supported by JSPS KAKENHI Grant Number 17J00631. H.-G.L. acknowledges financial support by the Sonderforschungsbereich SFB 881 ``The Milky Way System'' (subproject A4) of the German Research Foundation (DFG).
  J.M. acknowledges support from the ERC Consolidator Grant funding scheme (project STARKEY, G. A. n.615604). This work was partly funded by CNES.
\end{acknowledgements}

\appendix
\section{Tables of calibrated $\alpha$}

Table \ref{tab:alpha_m00} shows the calibrated $\alpha$ for all the models.

\begin{table*}
  \caption{Calibrated $\alpha$}
  \label{tab:alpha_m00}
  \centering
  \small
  \begin{tabular}{ccc|cc|cc|cc|cc|cc|cc}\hline\hline
    $T_{\rm eff}$ & $\log\,g$ & $\Delta s_{\rm 3D}$&\multicolumn{2}{c|}{CM} & \multicolumn{2}{c|}{CGM} 
    & \multicolumn{2}{c|}{CM2} & \multicolumn{2}{c|}{CGM2} & \multicolumn{2}{c|}{MLT(BV)} & \multicolumn{2}{c}{MLT(HVB)} \\
      & & & VAL-C  & Edd.   & VAL-C & Edd.   & VAL-C & Edd.   & VAL-C & Edd. & VAL-C & Edd.   & VAL-C & Edd.\\
    \hline
    4018 & 1.50 &  0.584 & 0.811&  0.764 & 0.618 & 0.583 & 0.317&$-$0.027&$-$0.017&$-$0.394& 1.834 & 1.741 & 1.930 & 1.830\\
    4476 & 2.50 &  0.342 & 0.906&  0.851 & 0.686 & 0.645 & 0.508&  0.368 & 0.174  &  0.038 & 1.919 & 1.816 & 2.054 & 1.941\\
    4477 & 4.00 &  0.068 & 1.168&  1.027 & 0.887 & 0.779 & 0.629&  0.231 & 0.231  &$-$0.178& 2.104 & 1.838 & 2.348 & 2.068\\
    4511 & 4.50 &  0.042 & 1.271&  1.053 & 0.966 & 0.801 & 0.541&  0.200 & 0.075  &$-$0.215& 2.210 & 1.737 & 2.486 & 2.024\\
    4586 & 3.20 &  0.181 & 1.002&  0.929 & 0.758 & 0.702 & 0.585&  0.392 & 0.240  & 0.048  & 1.980 & 1.847 & 2.161 & 2.013\\
    4775 & 3.20 &  0.227 & 0.985&  0.911 & 0.744 & 0.689 & 0.599&  0.464 & 0.251  & 0.136  & 1.977 & 1.844 & 2.146 & 1.999\\
    4969 & 2.50 &  0.622 & 0.868&  0.799 & 0.654 & 0.603 & 0.578&  0.480 & 0.219  & 0.144  & 1.889 & 1.758 & 1.997 & 1.855\\
    5037 & 3.00 &  0.392 & 0.914&  0.838 & 0.690 & 0.633 & 0.575&  0.451 & 0.224  & 0.134  & 1.915 & 1.775 & 2.051 & 1.897\\
    4924 & 3.50 &  0.195 & 1.016&  0.930 & 0.767 & 0.703 & 0.630&  0.495 & 0.279  & 0.169  & 1.996 & 1.844 & 2.179 & 2.010\\
    4955 & 4.00 &  0.117 & 1.106&  0.997 & 0.836 & 0.753 & 0.729&  0.550 & 0.360  & 0.215  & 2.063 & 1.872 & 2.282 & 2.070\\
    4981 & 4.50 &  0.071 & 1.183&  1.039 & 0.896 & 0.787 & 0.801&  0.567 & 0.433  & 0.229  & 2.122 & 1.865 & 2.365 & 2.083\\
    5432 & 3.50 &  0.354 & 0.953&  0.847 & 0.716 & 0.638 & 0.617&  0.473 & 0.255  & 0.158  & 1.940 & 1.751 & 2.091 & 1.881\\
    5476 & 4.00 &  0.209 & 1.048&  0.923 & 0.788 & 0.695 & 0.684&  0.525 & 0.317  & 0.203  & 2.024 & 1.807 & 2.216 & 1.972\\
    5488 & 4.50 &  0.121 & 1.137&  0.987 & 0.857 & 0.744 & 0.765&  0.577 & 0.394  & 0.249  & 2.090 & 1.832 & 2.318 & 2.029\\
    5775 & 4.44 &  0.176 & 1.093&  0.933 & 0.821 & 0.702 & 0.731&  0.533 & 0.355  & 0.216  & 2.051 & 1.780 & 2.261 & 1.956\\
    5885 & 3.50 &  0.579 & 0.951&  0.814 & 0.710 & 0.609 & 0.711&  0.532 & 0.305  & 0.191  & 1.948 & 1.706 & 2.085 & 1.817\\
    5927 & 4.00 &  0.340 & 1.023&  0.871 & 0.766 & 0.655 & 0.704&  0.523 & 0.316  & 0.194  & 2.018 & 1.756 & 2.190 & 1.896\\
    5861 & 4.50 &  0.182 & 1.099&  0.932 & 0.826 & 0.701 & 0.746&  0.541 & 0.361  & 0.218  & 2.059 & 1.775 & 2.271 & 1.951\\
    6102 & 4.25 &  0.324 & 1.021&  0.859 & 0.764 & 0.644 & 0.687&  0.502 & 0.306  & 0.180  & 1.981 & 1.706 & 2.158 & 1.848\\
    6176 & 3.50 &  0.778 & 0.985&  0.830 & 0.727 & 0.614 & 0.862&  0.648 & 0.399  & 0.263  & 1.970 & 1.705 & 2.106 & 1.811\\
    6233 & 4.50 &  0.270 & 1.085&  0.902 & 0.812 & 0.675 & 0.748&  0.540 & 0.354  & 0.213  & 2.056 & 1.751 & 2.256 & 1.909\\
    6432 & 4.25 &  0.472 & 1.012&  0.846 & 0.754 & 0.631 & 0.749&  0.541 & 0.332  & 0.205  & 1.963 & 1.684 & 2.131 & 1.816\\
    6486 & 4.00 &  0.660 & 0.998&  0.838 & 0.739 & 0.621 & 0.815&  0.610 & 0.372  & 0.246  & 1.943 & 1.676 & 2.097 & 1.796\\
    6458 & 4.50 &  0.349 & 1.074&  0.891 & 0.802 & 0.666 & 0.771&  0.556 & 0.361  & 0.222  & 2.044 & 1.740 & 2.236 & 1.891\\
    6725 & 4.25 &  0.659 & 1.027&  0.861 & 0.758 & 0.636 & 0.853&  0.647 & 0.400  & 0.269  & 1.948 & 1.681 & 2.115 & 1.811\\
    \hline
  \end{tabular}
  \tablefoot{The effective temperature $T_{\rm eff}$, surface gravity $g$ and entropy jump in the 3D models $\Delta s_{\rm 3D}$ are in unit of K, cm s$^{-2}$ and $10^9$ erg K$^{-1}$ g$^{-1}$, respectively.}
\end{table*}

\bibliographystyle{aa}
\bibliography{sonoi}

\begin{thebibliography}{47}
\expandafter\ifx\csname natexlab\endcsname\relax\def\natexlab#1{#1}\fi

\bibitem[{{Asplund}(2005)}]{Asplund05}
{Asplund}, M. 2005, \araa, 43, 481

\bibitem[{{B{\"o}hm-Vitense}(1958)}]{BV58}
{B{\"o}hm-Vitense}, E. 1958, \zap, 46, 108 (BV)

\bibitem[{{Broomhall} {et~al.}(2009){Broomhall}, {Chaplin}, {Davies},
  {Elsworth}, {Fletcher}, {Hale}, {Miller}, \& {New}}]{Broomhall09}
{Broomhall}, A.-M., {Chaplin}, W.~J., {Davies}, G.~R., {et~al.} 2009, \mnras,
  396, L100

\bibitem[{{Canuto}(1996)}]{Canuto96}
{Canuto}, V.~M. 1996, \apj, 467, 385

\bibitem[{{Canuto} {et~al.}(1996){Canuto}, {Goldman}, \& {Mazzitelli}}]{CGM96}
{Canuto}, V.~M., {Goldman}, I., \& {Mazzitelli}, I. 1996, \apj, 473, 550 (CGM)

\bibitem[{{Canuto} \& {Mazzitelli}(1991)}]{CM91}
{Canuto}, V.~M. \& {Mazzitelli}, I. 1991, \apj, 370, 295 (CM)

\bibitem[{{Christensen-Dalsgaard}(2008)}]{CD08}
{Christensen-Dalsgaard}, J. 2008, \apss, 316, 13

\bibitem[{{Dupret} {et~al.}(2002){Dupret}, {De Ridder}, {Neuforge}, {Aerts}, \&
  {Scuflaire}}]{Dupret02}
{Dupret}, M.-A., {De Ridder}, J., {Neuforge}, C., {Aerts}, C., \& {Scuflaire},
  R. 2002, \aap, 385, 563

\bibitem[{{Dupret} {et~al.}(2006){Dupret}, {Goupil}, {Samadi},
  {Grigahc{\`e}ne}, \& {Gabriel}}]{Dupret06}
{Dupret}, M.-A., {Goupil}, M.-J., {Samadi}, R., {Grigahc{\`e}ne}, A., \&
  {Gabriel}, M. 2006, in ESA Special Publication, Vol. 624, Proceedings of SOHO
  18/GONG 2006/HELAS I, Beyond the spherical Sun, 78.1

\bibitem[{{Freytag} {et~al.}(1999){Freytag}, {Ludwig}, \&
  {Steffen}}]{Freytag99}
{Freytag}, B., {Ludwig}, H.-G., \& {Steffen}, M. 1999, in Astronomical Society
  of the Pacific Conference Series, Vol. 173, Stellar Structure: Theory and
  Test of Connective Energy Transport, ed. A.~{Gimenez}, E.~F. {Guinan}, \&
  B.~{Montesinos}, 225

\bibitem[{{Freytag} {et~al.}(2002){Freytag}, {Steffen}, \& {Dorch}}]{Freytag02}
{Freytag}, B., {Steffen}, M., \& {Dorch}, B. 2002, Astronomische Nachrichten,
  323, 213

\bibitem[{{Freytag} {et~al.}(2012){Freytag}, {Steffen}, {Ludwig},
  {Wedemeyer-B{\"o}hm}, {Schaffenberger}, \& {Steiner}}]{Freytag12}
{Freytag}, B., {Steffen}, M., {Ludwig}, H.-G., {et~al.} 2012, Journal of
  Computational Physics, 231, 919

\bibitem[{{Galsgaard}(1996)}]{Galsgaard96}
{Galsgaard}, K. 1996, PhD thesis, Univ. Copenhagen

\bibitem[{{Gough} \& {Weiss}(1976)}]{Gough76}
{Gough}, D.~O. \& {Weiss}, N.~O. 1976, \mnras, 176, 589

\bibitem[{{Grevesse} \& {Noels}(1993)}]{GN93}
{Grevesse}, N. \& {Noels}, A. 1993, Physica Scripta Volume T, 47, 133

\bibitem[{{Grevesse} \& {Sauval}(1998)}]{Grevesse98}
{Grevesse}, N. \& {Sauval}, A.~J. 1998, \ssr, 85, 161

\bibitem[{{Grigahc{\`e}ne} {et~al.}(2005){Grigahc{\`e}ne}, {Dupret}, {Gabriel},
  {Garrido}, \& {Scuflaire}}]{Grigahcene05}
{Grigahc{\`e}ne}, A., {Dupret}, M.-A., {Gabriel}, M., {Garrido}, R., \&
  {Scuflaire}, R. 2005, \aap, 434, 1055

\bibitem[{{Gustafsson} {et~al.}(2008){Gustafsson}, {Edvardsson}, {Eriksson},
  {J{\o}rgensen}, {Nordlund}, \& {Plez}}]{Gustafsson08}
{Gustafsson}, B., {Edvardsson}, B., {Eriksson}, K., {et~al.} 2008, \aap, 486,
  951

\bibitem[{{Henyey} {et~al.}(1965){Henyey}, {Vardya}, \&
  {Bodenheimer}}]{Henyey65}
{Henyey}, L., {Vardya}, M.~S., \& {Bodenheimer}, P. 1965, \apj, 142, 841 (HVB)

\bibitem[{{Iglesias} \& {Rogers}(1996)}]{Iglesias96}
{Iglesias}, C.~A. \& {Rogers}, F.~J. 1996, \apj, 464, 943

\bibitem[{{Kritsuk} {et~al.}(2011){Kritsuk}, {Nordlund}, {Collins}, {Padoan},
  {Norman}, {Abel}, {Banerjee}, {Federrath}, {Flock}, {Lee}, {Li},
  {M{\"u}ller}, {Teyssier}, {Ustyugov}, {Vogel}, \& {Xu}}]{Kritsuk11}
{Kritsuk}, A.~G., {Nordlund}, {\AA}., {Collins}, D., {et~al.} 2011, \apj, 737,
  13

\bibitem[{{Ludwig} {et~al.}(2008){Ludwig}, {Caffau}, \& {Ku{\v
  c}inskas}}]{Ludwig08}
{Ludwig}, H.-G., {Caffau}, E., \& {Ku{\v c}inskas}, A. 2008, in IAU Symposium,
  Vol. 252, The Art of Modeling Stars in the 21st Century, ed. L.~{Deng} \&
  K.~L. {Chan}, 75--81

\bibitem[{{Ludwig} {et~al.}(2009){Ludwig}, {Caffau}, {Steffen}, {Freytag},
  {Bonifacio}, \& {Ku{\v c}inskas}}]{Ludwig09}
{Ludwig}, H.-G., {Caffau}, E., {Steffen}, M., {et~al.} 2009, \memsai, 80, 711

\bibitem[{{Ludwig} {et~al.}(1999){Ludwig}, {Freytag}, \& {Steffen}}]{Ludwig99}
{Ludwig}, H.-G., {Freytag}, B., \& {Steffen}, M. 1999, \aap, 346, 111 (LFS)

\bibitem[{{Magic} {et~al.}(2013){Magic}, {Collet}, {Asplund}, {Trampedach},
  {Hayek}, {Chiavassa}, {Stein}, \& {Nordlund}}]{Magic13}
{Magic}, Z., {Collet}, R., {Asplund}, M., {et~al.} 2013, \aap, 557, A26

\bibitem[{{Magic} {et~al.}(2015){Magic}, {Weiss}, \& {Asplund}}]{Magic15}
{Magic}, Z., {Weiss}, A., \& {Asplund}, M. 2015, \aap, 573, A89

\bibitem[{{Marques} {et~al.}(2013){Marques}, {Goupil}, {Lebreton}, {Talon},
  {Palacios}, {Belkacem}, {Ouazzani}, {Mosser}, {Moya}, {Morel}, {Pichon},
  {Mathis}, {Zahn}, {Turck-Chi{\`e}ze}, \& {Nghiem}}]{Marques13}
{Marques}, J.~P., {Goupil}, M.~J., {Lebreton}, Y., {et~al.} 2013, \aap, 549,
  A74

\bibitem[{{Montalb{\'a}n} {et~al.}(2004){Montalb{\'a}n}, {D'Antona}, {Kupka},
  \& {Heiter}}]{Montalban04}
{Montalb{\'a}n}, J., {D'Antona}, F., {Kupka}, F., \& {Heiter}, U. 2004, \aap,
  416, 1081

\bibitem[{{Montalb{\'a}n} {et~al.}(2001){Montalb{\'a}n}, {Kupka}, {D'Antona},
  \& {Schmidt}}]{Montalban01}
{Montalb{\'a}n}, J., {Kupka}, F., {D'Antona}, F., \& {Schmidt}, W. 2001, \aap,
  370, 982

\bibitem[{{Morel} {et~al.}(1994){Morel}, {van't Veer}, {Provost}, {Berthomieu},
  {Castelli}, {Cayrel}, {Goupil}, \& {Lebreton}}]{Morel94}
{Morel}, P., {van't Veer}, C., {Provost}, J., {et~al.} 1994, \aap, 286, 91

\bibitem[{{Mosumgaard} {et~al.}(2018){Mosumgaard}, {Ball}, {Silva Aguirre},
  {Weiss}, \& {Christensen-Dalsgaard}}]{Mosumgaard18}
{Mosumgaard}, J.~R., {Ball}, W.~H., {Silva Aguirre}, V., {Weiss}, A., \&
  {Christensen-Dalsgaard}, J. 2018, \mnras, 478, 5650

\bibitem[{{Rogers} \& {Nayfonov}(2002)}]{Rogers02}
{Rogers}, F.~J. \& {Nayfonov}, A. 2002, \apj, 576, 1064

\bibitem[{{Rosenthal} {et~al.}(1999){Rosenthal}, {Christensen-Dalsgaard},
  {Nordlund}, {Stein}, \& {Trampedach}}]{Rosenthal99}
{Rosenthal}, C.~S., {Christensen-Dalsgaard}, J., {Nordlund}, {\AA}., {Stein},
  R.~F., \& {Trampedach}, R. 1999, \aap, 351, 689

\bibitem[{{Salaris} \& {Cassisi}(2015)}]{Salaris15}
{Salaris}, M. \& {Cassisi}, S. 2015, \aap, 577, A60

\bibitem[{{Samadi} {et~al.}(2006){Samadi}, {Kupka}, {Goupil}, {Lebreton}, \&
  {van't Veer-Menneret}}]{Samadi06}
{Samadi}, R., {Kupka}, F., {Goupil}, M.~J., {Lebreton}, Y., \& {van't
  Veer-Menneret}, C. 2006, \aap, 445, 233

\bibitem[{{Sonoi} {et~al.}(2017){Sonoi}, {Belkacem}, {Dupret}, {Samadi},
  {Ludwig}, {Caffau}, \& {Mosser}}]{Sonoi17}
{Sonoi}, T., {Belkacem}, K., {Dupret}, M.-A., {et~al.} 2017, \aap, 600, A31

\bibitem[{{Sonoi} {et~al.}(2015){Sonoi}, {Samadi}, {Belkacem}, {Ludwig},
  {Caffau}, \& {Mosser}}]{Sonoi15}
{Sonoi}, T., {Samadi}, R., {Belkacem}, K., {et~al.} 2015, \aap, 583, A112

\bibitem[{{Steffen}(1993)}]{Steffen93}
{Steffen}, M. 1993, in Astronomical Society of the Pacific Conference Series,
  Vol.~40, IAU Colloq. 137: Inside the Stars, ed. W.~W. {Weiss} \& A.~{Baglin},
  300

\bibitem[{{Stein} \& {Nordlund}(1998)}]{Stein98}
{Stein}, R.~F. \& {Nordlund}, {\AA}. 1998, \apj, 499, 914

\bibitem[{{Trampedach} {et~al.}(2013){Trampedach}, {Asplund}, {Collet},
  {Nordlund}, \& {Stein}}]{Trampedach13}
{Trampedach}, R., {Asplund}, M., {Collet}, R., {Nordlund}, {\AA}., \& {Stein},
  R.~F. 2013, \apj, 769, 18

\bibitem[{{Trampedach} {et~al.}(1997){Trampedach}, {Christensen-Dalsgaard},
  {Nordlund}, \& {Stein}}]{Trampedach97}
{Trampedach}, R., {Christensen-Dalsgaard}, J., {Nordlund}, A., \& {Stein}, R.
  1997, in Astrophysics and Space Science Library, Vol. 225, SCORe'96 : Solar
  Convection and Oscillations and their Relationship, ed. F.~P. {Pijpers},
  J.~{Christensen-Dalsgaard}, \& C.~S. {Rosenthal}, 73--76

\bibitem[{{Trampedach} {et~al.}(2014{\natexlab{a}}){Trampedach}, {Stein},
  {Christensen-Dalsgaard}, {Nordlund}, \& {Asplund}}]{Trampedach14a}
{Trampedach}, R., {Stein}, R.~F., {Christensen-Dalsgaard}, J., {Nordlund},
  {\AA}., \& {Asplund}, M. 2014{\natexlab{a}}, \mnras, 442, 805

\bibitem[{{Trampedach} {et~al.}(2014{\natexlab{b}}){Trampedach}, {Stein},
  {Christensen-Dalsgaard}, {Nordlund}, \& {Asplund}}]{Trampedach14b}
{Trampedach}, R., {Stein}, R.~F., {Christensen-Dalsgaard}, J., {Nordlund},
  {\AA}., \& {Asplund}, M. 2014{\natexlab{b}}, \mnras, 445, 4366

\bibitem[{{Ventura} {et~al.}(2008){Ventura}, {D'Antona}, \&
  {Mazzitelli}}]{Ventura08}
{Ventura}, P., {D'Antona}, F., \& {Mazzitelli}, I. 2008, \apss, 316, 93

\bibitem[{{Vernazza} {et~al.}(1981){Vernazza}, {Avrett}, \&
  {Loeser}}]{Vernazza81}
{Vernazza}, J.~E., {Avrett}, E.~H., \& {Loeser}, R. 1981, \apjs, 45, 635

\bibitem[{{Wedemeyer} {et~al.}(2004){Wedemeyer}, {Freytag}, {Steffen},
  {Ludwig}, \& {Holweger}}]{Wedemeyer04}
{Wedemeyer}, S., {Freytag}, B., {Steffen}, M., {Ludwig}, H.-G., \& {Holweger},
  H. 2004, \aap, 414, 1121

\bibitem[{{Wolf}(1983)}]{Wolf83}
{Wolf}, B.~E. 1983, \aap, 127, 93

\end{thebibliography}

\end{document}